\documentclass[10pt,preprint]{aastex}

\usepackage{amsmath,natbib}

\begin{document}

\title{Hydrodynamic Simulations of Tilted Thick-Disk Accretion onto a Kerr Black Hole}

\author{P. Chris Fragile and Peter Anninos}
\affil{University of California,
Lawrence Livermore National Laboratory, Livermore, CA 94550}


\begin{abstract}
We present results from fully general relativistic three-dimensional
numerical studies of thick-disk accretion onto a rapidly-rotating
(Kerr) black hole with a spin axis that is tilted (not aligned) with
the angular momentum vector of the disk. We initialize the problem
with the solution for an aligned, constant angular momentum,
accreting thick disk, which is then allowed to respond to the
Lense-Thirring precession of the tilted black hole. The precession
causes the disk to warp, beginning at the inner edge and moving out
on roughly the Lense-Thirring precession timescale. The propagation
of the warp stops at a radius in the disk at which other dynamical
timescales, primarily the azimuthal sound-crossing time, become
shorter than the precession time. At this point, the warp
effectively freezes into the disk and the evolution becomes
quasi-static, except in cases where the sound-crossing time in the
bulk of the disk is shorter than the local precession timescale. We
see evidence that such disks undergo near solid-body precession
after the initial warping has frozen in. Simultaneous to the warping
of the disk, there is also a tendency for the midplane to align with
the symmetry plane of the black hole due to the preferential accretion 
of the most tilted disk gas. This alignment is not as
pronounced, however, as it would be if more
efficient angular momentum transport (e.g. from viscosity or
magneto-rotational instability) were considered.
\end{abstract}


\section{Introduction}
\label{sec:introduction}
Quasars, active galactic nuclei (AGN), X-ray binaries,
core-collapse supernovas, and gamma-ray bursts (GRBs)
are among the most luminous
objects in the universe and are therefore of considerable observational
interest.  They are also
all likely to contain a central, strongly gravitating,
compact object and be powered by some form of
accretion.  They are therefore also of considerable
theoretical interest.  However, the nonlinear, time-dependent,
multi-dimensional nature of the physical laws governing these objects
makes a study of them from first principles difficult.  While such
studies can provide critical insight, there is strong motivation
to apply a numerical
approach.

Many numerical simulations have been carried out over the past three
decades studying the time-dependent evolution of accretion flows
around black holes in the hydrodynamic
\citep[e.g.][]{wil72,haw84b,haw91} and MHD
\citep{koi99,gam03a,dev03b} regimes. However, due to the complex
nature of this problem, a variety of simplifying assumptions are
normally implemented. A common simplification in much of the early
work was to assume a non-rotating (Schwarzschild) black hole, often
approximated using a pseudo-Newtonian \citep{pac80} potential.  A
notable exception was the work of \citet{wil72}, which considered
the spherical infall of material with a non-zero specific angular
momentum toward a Kerr black hole using the full metric, although
restricted to two spatial dimensions. Although the assumption of a
non-rotating black hole considerably simplifies the problem while
still allowing useful insight, this does not appear to be well
represented among observed black-hole systems
\citep[e.g.][]{elv02,yu02}, nor does it agree with studies of black
hole spin evolution \citep{gam04,vol04} which suggest that most
black holes are likely to be spinning rapidly.

The spin of a black hole (often denoted by the specific angular
momentum $a = J/M$) can have important consequences on the accretion
rate and the geometry of the accretion flow. One obvious example is
the dependence of $r_{ms}$, the radius of the marginally stable
circular orbit, on the spin parameter [$r_{ms}=6r_G$ for $a=0$
(non-rotating black hole); $r_{ms}=r_G$ for $a=+M$ (maximal prograde
rotation); $r_{ms}=9r_G$ for $a=-M$ (maximal retrograde rotation),
where $r_G=GM/c^2$ is the gravitational radius \citep{bar72}].
Another example is the reversal of flow direction in the vicinity of
a rotating black hole for a retrograde accretion disk \citep[see
Figure 6 of][]{dev02}.

Another common simplification in black hole accretion models has
been the assumption that the midplane of the accretion flow is
aligned with the symmetry plane of the black hole. This has
historically been necessitated in numerical work by limitations in
computing resources which have often limited simulations to
two-dimensional axisymmetry. The justification for this treatment
often relies on the Bardeen-Petterson effect \citep{bar75} through
which a spinning black hole may align the inner region of the
accretion flow with the symmetry plane of the black hole. However,
now that computational resources are available to perform
three-dimensional simulations of black-hole accretion, there are
many compelling reasons to revisit the question of black-hole - disk
alignment. First, the formation avenues for many black-hole - disk
systems favor, or at least allow for, an initially tilted
configuration (see \citet{fra01a} for a review of the arguments).
Second, although the Bardeen-Petterson effect may align the inner
disk, this alignment may not extend nearly as far out as was
originally expected [$15-30 r_G$ \citep{nel00} rather than $100-1000
r_G$].  Many numerical simulations extend to larger radii than this
and, therefore, may not be justified in assuming that the midplane
of the disk is aligned with the symmetry plane of the black hole.
Finally, although there is a residual torque between the tilted
outer disk and black hole which will eventually completely align the
system \citep{ree78,sch96,nat98}, this happens over such a long
timescale that these systems remain tilted for most of their
observational lifetimes.


If an accretion disk is misaligned or tilted, it will be subject
to differential Lense-Thirring precession.  This is
ultimately what gives rise to the Bardeen-Petterson effect.
For an ideal test particle in
a slightly tilted orbit at a radius $r$ around a black hole of mass $M$ and
specific angular momentum $a$,
gravitomagnetic torques cause the
orbit to precess at a frequency $\Omega_{LT}=2aM/r^3$.  Close to the
black hole, this is comparable to the orbital frequency
[$\Omega_{Kep}=(M/r^3)^{1/2}$].  However, because of its
strong radial dependence, Lense-Thirring precession
becomes much weaker far from the hole.  Nevertheless, since
it has a cumulative effect, it can build up appreciably over a
sufficiently large number of orbits.  In an
accretion disk, this cumulative build-up
is limited by dynamic responses within the disk.
For a steady-state disk, Lense-Thirring precession is expected
to be important out to a unique, nearly constant transition
radius \citep{bar75,kum85}.

A comprehensive understanding of tilted accretion disks can be
aided by the insights of direct
numerical simulations.  An important first step
was taken by \citet{nel00}, who utilized a
three-dimensional Newtonian SPH code to study tilted thin
accretion disks subject to a post-Newtonian gravitomagnetic
torque.  In the current work we present results from
three-dimensional, fully relativistic simulations of tilted
black-hole accretion flows.  A general relativistic
treatment ensures that all important relativistic features,
such as the cusp in the potential and the Einstein and Lense-Thirring
precessions of the orbits, as well as any higher order couplings
of these features, are treated properly.
In this first study, we consider
relatively simple, inviscid, thick accretion disks.
These disks are well studied in the literature and provide a
convenient starting point for our work.
We should mention, however, that effective angular momentum transport
(such as through viscous coupling as considered in \citet{nel00}),
would likely have a considerable impact on our results.

Unless otherwise stated, we use geometrized units ($G=c=1$)
throughout this paper.  Units of length are parameterized in terms
of the gravitational radius of the black hole, $r_G=GM/c^2$.
We also adopt
the standard convention in which Greek (Latin)
indices refer to 4(3)-space components.

\section{Inviscid Accreting Thick Disks}

Geometrically thick disks (often referred to as
black hole tori or toroidal neutron stars) have been proposed to form
through a number of scenarios, including
the collapse of supermassive neutron stars \citep{vie98},
the iron-core collapse of a massive stars \citep{mac99},
and the mergers of black hole - neutron star binaries \citep{lee99}. 
They may also exist in the inner regions of more standard accretion 
flows. 
These disks are convenient to study numerically as they
represent an equilibrium solution in the limit that self-gravity,
radiation, and viscosity are negligible.
However, they are subject to numerous instabilities:
1) the so-called ``runaway instability'' in tori with non-negligible
mass and accretion \citep{fon02};
2) the magneto-rotational instability \citep[MRI,][]{bal91}; and
3) the Papaloizou-Pringle instability \citep[PPI,][]{pap84}.
In this work, we assume that the mass of the torus is negligible
compared to the mass of the black hole, thus allowing us to
neglect the runaway instability.  This also allows us to
ignore the self-gravity of the torus gas and treat the background
metric as fixed.  We also ignore magnetic fields in the current
work, so the MRI is not applicable.  Below we discuss how the PPI
affects our work in more detail.

The analytic solution for the structure of a thick disk has been
worked out for both non-rotating and rotating black holes in the
limit of axisymmetry \citep{fis76,abr78}.  However, no such solution
has been worked out for more general tilted flows. Furthermore, it
is not clear that a steady-state solution would result.  In light of
the violent formation scenarios proposed above, however, it seems likely a
newly formed torus will be highly disturbed and possibly have an
orbital plane that is tilted with respect to the symmetry plane of
the black hole. This, in fact, is part of the motivation for this
numerical study. Nevertheless, we are left with a decision about how
to initialize the problem.  We choose to initialize all our disks
with the analytic solution for an axisymmetric thick disk, which is
subsequently allowed to respond to the Lense-Thirring precession of
a tilted black hole.

For untilted accreting thick disks, the gas flows in an
effective (gravitational plus centrifugal) potential.
Since axisymmetric thick disks have been described in
great detail elsewhere \citep[e.g.][]{koz78}, we present only a brief
discussion here.
Surfaces of constant pressure in the torus
are determined from the relativistic analog of the Newtonian effective
potential $\Phi$,
\begin{equation}
\Phi-\Phi_{in} = -\int^P_0 \frac{\mathrm{d}P}{\rho h} ~,
\end{equation}
where $P$ is the fluid pressure, $\rho$ is the density, $h$ is the
relativistic enthalpy, and $\Phi_{in}$ is the potential at the
boundary of the torus (where $P=0$).  For constant angular momentum
$l$, the form of the potential reduces to $\Phi=\ln(-u_t)$, where
$u_t$ is the specific binding energy.  Provided $l>l_{ms}$, where
$l_{ms}$ is the angular momentum of the marginally stable Keplerian
orbit, the potential $\Phi(r,\theta)$ will have a saddle point
$\Phi_{cusp}$ at $r=r_{cusp}$, $\theta=\pi/2$.  We can define the
parameter $\Delta \Phi = \Phi_{in} - \Phi_{cusp}$ as the potential
barrier (energy gap) at the inner edge of the torus. If $\Delta \Phi
<0$, the torus lies entirely within its Roche lobe. Such a
configuration is marginally stable with respect to local
axisymmetric perturbations and unstable to low-order
non-axisymmetric PPI modes \citep{pap84}.  If $\Delta \Phi > 0$, the
torus overflows its Roche lobe and accretion occurs through
pressure-gradient forces across the cusp. This accretion occurs
without dissipation of angular momentum; thus viscosity is not
required for accretion.  This accretion generally suppresses the
growth of instabilities in the torus \citep{haw91}.  All of the
simulations presented in this work begin with $\Delta \Phi > 0$.

\section{Hydrodynamic Equations in Kerr Geometry}

In this work, we solve
the system of equations for relativistic hydrodynamics
in flux-conserving form \citep{ann03a}
\begin{eqnarray}
 \frac{\partial D}{\partial t} + \frac{\partial (DV^i)}{\partial x^i} &=& 0 ,
      \label{eqn:av_de} \\
 \frac{\partial E}{\partial t} + \frac{\partial (EV^i)}{\partial x^i}
 + P\frac{\partial W}{\partial t} + P\frac{\partial (WV^i)}{\partial x^i} &=& 0
      \label{eqn:av_en} \\
 \frac{\partial S_j}{\partial t} + \frac{\partial (S_j V^i)}{\partial x^i}
 - \frac{S^\mu S^\nu}{2S^0} \frac{\partial g_{\mu\nu}}{\partial x^j}
 + \sqrt{-g} \frac{\partial P}{\partial x^j} &=& 0 ,
      \label{eqn:av_mom}
\end{eqnarray}
where $g$ is the determinant of the 4-metric,
$W=\sqrt{-g} u^t$ is the relativistic boost factor,
$D=W\rho$ is the generalized fluid density,
$P=(\Gamma-1)E/W$ is the fluid pressure,
$V^i=u^i/u^t$ is the transport velocity,
$S_i = W\rho h u_i$ is the covariant momentum density,
and $E=We=W\rho\epsilon$ is the generalized internal energy density.

These equations are evolved in
a ``tilted'' Kerr-Schild polar coordinate system
$({t},{r},{\vartheta},{\varphi})$.
This coordinate system is related
to the usual (untilted) Kerr-Schild coordinates
$({t},{r},{\theta},{\phi})$
through a simple rotation
about the ${y}$-axis by an angle $\beta_0$, such that
\begin{equation}
\left( \begin{array}{c} \sin{{\vartheta}}\cos{{\varphi}} \\
                        \sin{{\vartheta}}\sin{{\varphi}} \\
                        \cos{{\vartheta}} \end{array} \right)
 = \left( \begin{array}{ccc} \cos{\beta_0} & 0 & -\sin{\beta_0} \\
                                 0 & 1 & 0 \\
                  \sin{\beta_0} & 0 & \cos{\beta_0} \end{array} \right)
       \left( \begin{array}{c} \sin{{\theta}}\cos{{\phi}} \\
                        \sin{{\theta}}\sin{{\phi}} \\
                        \cos{{\theta}} \end{array} \right) ~.
\label{eqn:tiltarray}
\end{equation}
The Kerr line element in the tilted $({t},{r},{\vartheta},{\varphi})$
coordinate system is
\begin{eqnarray}
\mathrm{d}s^2 & = & - \left(1-\frac{2Mr}{\varrho^2}\right)\mathrm{d}t^2
+ \frac{4Mr}{\varrho^2}\mathrm{d}t\mathrm{d}r
- \frac{4Mar\sin^2\theta}{\varrho^2}P_1 \mathrm{d}t\mathrm{d}\vartheta \nonumber \\
 & & - \frac{4Mar\sin^2\theta}{\varrho^2}P_2 \mathrm{d}t\mathrm{d}\varphi
+ \left(1 + \frac{2Mr}{\varrho^2}\right) \mathrm{d}r^2 \nonumber \\
 & & - 2a\sin^2\theta\left(1 - \frac{2Mr}{\varrho^2}\right) P_1 \mathrm{d}r\mathrm{d}\vartheta \nonumber \\
 & & - 2a\sin^2\theta\left(1 - \frac{2Mr}{\varrho^2}\right) P_2 \mathrm{d}r\mathrm{d}\varphi \nonumber \\
 & & + \left(\varrho^2 T_1^2 + A P_1^2 \right) \mathrm{d}\vartheta^2
+ 2\left(\varrho^2 T_1 T_2 + A P_1 P_2 \right) \mathrm{d}\vartheta \mathrm{d}\varphi \nonumber \\
 & & + \left(\varrho^2 T_2^2 + A P_2^2 \right) \mathrm{d}\varphi^2 ~,
\end{eqnarray}
with
\begin{eqnarray}
\varrho^2 & \equiv & r^2+a^2\cos^2\theta \nonumber \\
\Delta & \equiv & r^2 - 2Mr + a^2 \nonumber \\
A & \equiv & \frac{\left[\left(r^2+a^2\right)^2 - \Delta a^2 \sin^2 \theta\right]\sin^2\theta}{\varrho^2} \nonumber \\
T_1 & \equiv & \frac{-1}{\sin\theta}\frac{\mathrm{d}(\cos\theta)}{\mathrm{d}\vartheta} \nonumber \\
T_2 & \equiv & \frac{-1}{\sin\theta}\frac{\mathrm{d}(\cos\theta)}{\mathrm{d}\varphi} \nonumber \\
P_1 & \equiv & \frac{1}{\cos\phi}\frac{\mathrm{d}(\sin\phi)}{\mathrm{d}\vartheta} \nonumber \\
P_2 & \equiv & \frac{1}{\cos\phi}\frac{\mathrm{d}(\sin\phi)}{\mathrm{d}\varphi} ~,
\end{eqnarray}
where $\sin\theta$, $\cos\theta$, $\sin\phi$, and $\cos\phi$ are given by equation \ref{eqn:tiltarray}.
For a Schwarzschild ($a=0$) or an aligned Kerr black hole
($\beta_0=n\pi$, where $n=0,1,2,$...), the
$({t},{r},{\vartheta},{\varphi})$ and
$({t},{r},{\theta},{\phi})$ frames are equivalent.

The computational advantages of the ``horizon-adapted''
Kerr-Schild form of the
Kerr metric are described in \citet{pap98} and \citet{fon98b}.
The primary advantage is that, unlike Boyer-Lindquist coordinates,
there are no singularities in the metric terms at the event
horizon.  This is particularly important for numerical calculations
as it allows one to place the grid boundaries inside the horizon,
thus ensuring that they are causally disconnected from the
rest of the flow.

\section{Numerical Methods}
\label{sec:methods}

Our main purpose in this work is to examine the structure
of tilted accreting thick disks orbiting around a Kerr black hole.
In practice, we actually initialize the disks
aligned with the computational grid
and tilt the black hole through a transformation of the metric.
We have tested problems with the disk
tilted relative to the grid and found similar results.
We carry out these simulations using Cosmos, a massively parallel,
multidimensional, radiation-chemo-magnetohydrodynamic code for both Newtonian
and relativistic flows.
The relativistic capabilities and tests of Cosmos are
discussed in \citet{ann03a}.  The present work utilizes the
zone-centered artificial viscosity hydrodynamics package in Cosmos.

The simulations are initialized with the analytic solution for
an axisymmetric torus with constant specific angular momentum
$l$ around a rotating black hole with spin $a$.
For the initialization, we assume an isentropic equation of state
$P=\kappa \rho^\Gamma$,
although during the evolution the adiabatic form
$P=(\Gamma-1)E/W$ is used to recover the pressure when solving
equations (\ref{eqn:av_de}) - (\ref{eqn:av_mom}).
For all but one model, we set $\Gamma=4/3$ and $\kappa=1.06 \times 10^{11}$ 
(in cgs units).  
The surface of the torus is defined by the effective potential at
the boundary of the torus $\Phi_{in}$, although in practice the torus
is cut off where the fluid density and pressure match the background
values.
All of our runs are initialized with a positive energy gap
$\Delta \Phi= \Phi_{in} - \Phi_{cusp} > 0$,
where $\Phi_{cusp}$ is the saddle point in the potential at
$r=r_{cusp}$, $\theta=\pi/2$.  With this choice, a small fraction of
the torus mass is initially out of equilibrium.

\begin{figure}
\figurenum{1}
\plotone{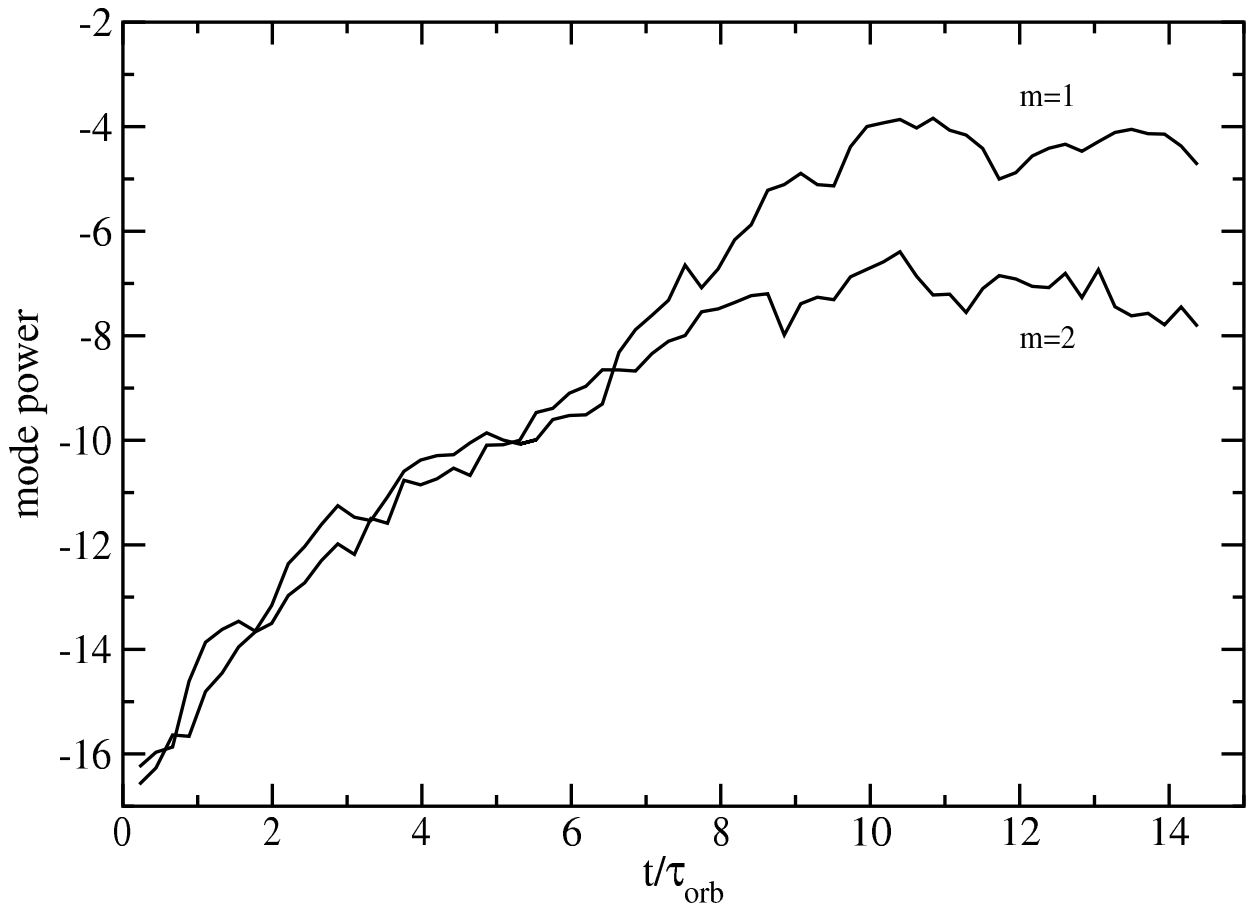}
\caption{PPI mode growth for comparison with Model A3p of \citet{dev02}.}
\label{fig:ppi}
\end{figure}

The simulations are carried out on a grid extending from
$0.98r_{BH}\le r \le r_{max}$, $0 \le \vartheta \le \pi$, and $0 \le
\varphi < 2\pi$, where $r_{BH}$ is the radius of the black hole
horizon. We choose $r_{max}$ such that the disk is initially
contained entirely within the computational domain. In order to
increase the resolution in the inner region of the disk, we replace
the radial coordinate ${r}$ with a logarithmic coordinate $\eta = 1
+ \ln (r/r_{BH})$.  In terms of this logarithmic coordinate, the
grid extends from $0.98 \le \eta \le \eta_{max}$, with a typical
radial grid spacing of $0.2 r_G$ near the horizon and $6r_G$ near
the outer boundary.  Zones are evenly spaced in both angular
dimensions.  The grid is normally resolved with $96\times64\times96$
zones along the dimensions $\eta$, $\vartheta$, $\varphi$. This
resolution is sufficient to resolve key disk properties such as the
PPI.  To illustrate this point and demonstrate the ability of our
code to reproduce other published work, we plot in Figure
\ref{fig:ppi} the growth of the $m=1$ and $m=2$ PPI modes for a
simulation comparable to Model A3p of \citet{dev02} using a
$64\times64\times64$ grid. Our results are very similar to those of
de Villiers \& Hawley as seen by comparing to their Figure 7.
Additional verification of our results is provided by comparing
simulations at different grid resolutions in order to study the
self-convergence of our solutions.

In the ``background'' regions not determined by the initial
torus solution, we initialize the gas following the spherical Bondi
accretion solution.  We fix the parameters of this solution such that
the rest mass present in the background is negligible
($\lesssim 10^{-2}$) compared to
the mass in the torus.  Similarly, the mass accretion from the background
is generally small ($\lesssim 10^{-3}$)
compared to the mass accretion from the disk.
The outer radial boundary
is held fixed with the analytic Bondi solution for all evolved
fields.  The inner radial
boundary uses simple flat (zero gradient) boundary conditions.
Data are shared appropriately across angular boundaries.

After the torus and background are initialized, the black hole is
tilted by an angle $\beta_0$ through a transformation of the metric.
Each model is evolved to
$t \gtrsim 10 \tau_{orb}$, where
$\tau_{orb}$ is the orbital period measured at the pressure (and density) maximum
of the axisymmetric torus (denoted $r_{center}$).  Most of the disks
achieve a quasi-steady state
[i.e. variations in measured disk properties,
such as the twist and tilt, are small ($<10$\%)
over dynamical timescales] by the end of the simulations. The principle exceptions 
are disks that undergo near solid-body precession and disks that 
accrete a large percentage of their mass ($\gtrsim 50\%$)
onto the black hole.

To describe our results more thoroughly, we track a
number of quantities throughout the simulations, including the
mass accretion rate, the PPI mode growth for the
azimuthal wavenumbers $m=1$ and 2, and
the precession angle (twist) and angle of inclination (tilt)
of the disk.
The mass accretion rate is defined as
\begin{equation}
\dot{M}(r) = \int^{2\pi}_0 \int^\pi_0 D V^r \mathrm{d}\vartheta \mathrm{d}\varphi ~.
\end{equation}
The accretion rates are converted to units with a
polytropic constant $\kappa=1.5 \times 10^{20}$ (cgs units) and
mass $M=1M_\odot$, consistent with \citet{igu97}.
This value is then normalized by the
Eddington accretion rate
$\dot{M}_{Edd}=L_{Edd}/c^2\approx1.4 \times 10^{17} (M/M_\odot)$
g s$^{-1}$.  The conversion of $\dot{M}$ is accomplished
through the relation
\begin{equation}
\left(\frac{\dot{M}}{\dot{M}_{Edd}}\right)_{new} = \frac{(M)_{new}}
{(M)_{old}} \left( \frac{\kappa_{new}}{\kappa_{old}}
\right)^{-1/(\Gamma-1)} \left(\frac{\dot{M}}{\dot{M}_{Edd}}\right)_{old} ~.
\end{equation}
Using this common scaling allows for easy comparison of our results.
Nevertheless, we caution the reader against attaching any 
particular physical significance to these values since the mass 
accretion rates and densities in these simulations are essentially 
arbitrary.

The $m=1$ and 2 Fourier modes of the PPI are extracted by computing
azimuthal density averages of the following form \citep{dev02}
\begin{eqnarray}
\mathrm{Re}[k_m(r)] = \int_0^{2\pi} \rho (r,\pi/2,\varphi)
            \cos(m\varphi) d\varphi ~, \\
\mathrm{Im}[k_m(r)] = \int_0^{2\pi} \rho (r,\pi/2,\varphi)
            \sin(m\varphi) d\varphi ~.
\end{eqnarray}
The power in mode $m$ is then
\begin{equation}
f_m = \frac{1}{r_{max}-r_{min}}\int_{r_{min}}^{r_{max}} \ln
    (\{\mathrm{Re}[k_m(r)]\}^2 + \{\mathrm{Im}[k_m(r)]\}^2) dr ~,
\end{equation}
where $r_{min}$ and $r_{max}$ are the approximate inner and outer
edges of the disk.

We define the precession angle (twist) as \citep{nel00}
\begin{equation}
\gamma(r) = \arccos\left[ \frac{\mathbf{J}_{BH} \times \mathbf{J}_{Disk}(r)}
{\vert \mathbf{J}_{BH} \times \mathbf{J}_{Disk}(r) \vert}
\cdot \hat{y}\right] ~,
\label{eq:twist}
\end{equation}
where
\begin{equation}
\mathbf{J}_{BH} = \left( -a M \sin\beta_0 \hat{x}, 0, a M
\cos\beta_0 \hat{z} \right)
\end{equation}
is the angular momentum vector of the black hole and
\begin{equation}
\mathbf{J}_{Disk}(r) = \left[ (J_{Disk})_1 \hat{x}, (J_{Disk})_2
\hat{y}, (J_{Disk})_3 \hat{z} \right]
\end{equation}
is the angular momentum vector of the disk in an asymptotically flat
space, where
\begin{equation}
(J_{Disk})_\rho = \frac{\epsilon_{\mu \nu \sigma \rho}
                         L^{\mu \nu} S^\sigma}
                       {2 \sqrt{-S^\alpha S_\alpha}}
\end{equation}
and
\begin{equation}
L^{\mu \nu} = \int \left( x^\mu T^{\nu 0} - x^\nu T^{\mu 0} \right) \mathrm{d}^3 x ,
\end{equation}
with $S^\sigma = \int T^{\sigma 0} \mathrm{d}^3 x$, $T^{\mu\nu} =
\rho h u^\mu u^\nu + Pg^{\mu\nu}$, and $\beta_0$ is the initial tilt
between the angular momentum vectors of the black hole and the disk.
The equations for $L^{\mu \nu}$ and $S^\sigma$ are integrated over
concentric radial shells of the grid. The unit vector $\hat{y}$
points along the axis about which the black hole is initially tilted
and $\hat{z}$ points along the initial symmetry axis of the disk.
Thus, from Equation (\ref{eq:twist}), $\gamma(r) = 0$ throughout the
disk at $t=0$. In order to capture twists larger than $180^\circ$,
we also track the projection of $\mathbf{J}_{BH} \times
\mathbf{J}_{Disk}(r)$ onto $\hat{x}$, allowing us to break the
degeneracy in $\arccos$.
We use the condition $\gamma=1$ to define a transition radius $r_T$
between an inner twisted disk (precession greater than 1 radian)
and the outer undisturbed disk. 

The angle of inclination (tilt) in the disk is defined as \citep{nel00}
\begin{equation}
\beta(r) = \arccos\left[ \frac{\mathbf{J}_{BH} \cdot \mathbf{J}_{Disk}(r)}
{\vert\mathbf{J}_{BH} \vert \vert\mathbf{J}_{Disk}(r) \vert} \right] ~.
\end{equation}
The models are set up such that, initially, $\beta(r)=\beta_0$ throughout
the disk.
For tilted disks ($\beta_0 \ne 0$) such as the ones considered in this work,
Lense-Thirring precession proceeds most rapidly
closest to the black hole and gradually affects the disk further out
as the simulations progress.

We expect differential Lense-Thirring precession to dominate whenever 
the precession timescale $\tau_{LT} = \Omega_{LT}^{-1}$ is shorter 
than local dynamical timescales in the disk \citep{bar75,kum85}.  
We consider two possible limiting timescales:
the mass accretion timescale $\tau_{acc} =
(r-r_{BH})/\overline{V}^r$, where $\overline{V}^r$ is the average
inflow velocity of the torus gas, and the azimuthal sound-crossing
time defined as $\tau_{cs}=\pi r/c_s$, where $c_s$ is the local
sound speed in the density midplane. In this work we determine the
accretion timescale numerically from
$\overline{V}^r=\dot{M}/\lambda$, where $\lambda(r) \equiv \int \int
D \mathrm{d}\vartheta \mathrm{d}\varphi$.  
Initially the sound speed in the disk is given by 
$c_s^2 = (\Gamma-1)(1-u_t/u_{t,in})$.  During the evolution, 
the local sound speed is
recovered from the fluid state through the relation $c_s^2=\Gamma
(\Gamma-1)P/[(\Gamma-1)\rho + \Gamma P]$.  Since the sound speed is
defined in the proper frame of the fluid, it is not strictly
accurate to compare $\tau_{cs}$ to quantities defined using the
coordinate time (such as $\tau_{LT}$).  However, the corrections are
only of order $Mr/(r^2+a^2)$ and should therefore be small ($\lesssim
20$\%) for the radii of interest here.

\section{Results}
\label{sec:results}

In this work we present results from eleven different models, varying
the spin of the black hole, the angular momentum and tilt of the
torus, the magnitude of the energy gap, the sound speed within the 
disk, and the size and resolution
of the grid. The basic parameters for each model are listed in Table
\ref{tab:models}. Many of our parameters closely match those of
Models II and III in \citet{igu97}, thus providing us with an avenue
to compare our results with other numerical simulations.  However,
we emphasize that there are a number of significant differences
between their simulations and ours: all of the simulations reported
here are performed in three dimensions and all but one are tilted;
we use the Kerr-Schild form of the Kerr metric; we use a logarithmic
radial coordinate for higher resolution near the black hole; and we
choose a different inner surface potential $\Phi_{in}$. In the
following sections we describe the results of each of our models in
more detail.

\begin{deluxetable}{ccccccc}
\tablewidth{0pt}
\tablecaption{Tilted Thick-Disk Models \label{tab:models}}
\tablehead{
\colhead{ } &
\colhead{Grid}  &
\colhead{ } &
\colhead{ } &
\colhead{ } &
\colhead{ } &
\colhead{ }  \\
\colhead{Model} &
\colhead{($N_\eta\times N_\vartheta\times N_\varphi$)}  &
\colhead{$a/M$} &
\colhead{$l/M$} &
\colhead{($t_{orb}/M$)\tablenotemark{a}} &
\colhead{$\Delta \Phi$} &
\colhead{$\beta_0$}
}
\startdata
t90    & $96\times64\times96$ & 0.9 & 2.6088 & 46.1 & 0.03  & 0           \\ 
t915XL & $24\times24\times24$ & 0.9 & 2.6088 & 46.1 & 0.03  & 15$^\circ$  \\ 
t915L  & $48\times24\times48$ & 0.9 & 2.6088 & 46.1 & 0.03  & 15$^\circ$  \\ 
t915H  & $96\times64\times96$ & 0.9 & 2.6088 & 46.1 & 0.03  & 15$^\circ$  \\ 
t915H2 & $96\times64\times96$ & 0.9 & 2.6088 & 46.1 & 0.04  & 15$^\circ$  \\ 
t930L  & $48\times24\times48$ & 0.9 & 2.6088 & 46.1 & 0.03  & 30$^\circ$  \\ 
t930H  & $96\times64\times96$ & 0.9 & 2.6088 & 46.1 & 0.03  & 30$^\circ$  \\ 
t915$\Gamma$\tablenotemark{b} & $48\times24\times48$ & 0.9 & 2.6088 & 46.1 & 0.03  & 15$^\circ$  \\ 
t915R  & $96\times64\times96$ & -0.9\tablenotemark{c} & 4.4751 & 255. & 0.0175 & 15$^\circ$  \\ 
t515L  & $48\times24\times48$ & 0.5 & 3.385  & 118. & 0.004 & 15$^\circ$  \\ 
t515H  & $96\times64\times96$ & 0.5 & 3.385  & 118. & 0.004 & 15$^\circ$     
\enddata
\tablenotetext{a}{Orbital period measured at the pressure center of
the torus.}
\tablenotetext{b}{$\Gamma=5/3$.}
\tablenotetext{c}{Retrograde torus.}
\end{deluxetable}

\subsection{Model t90}
\begin{figure}
\figurenum{2a}
\plotone{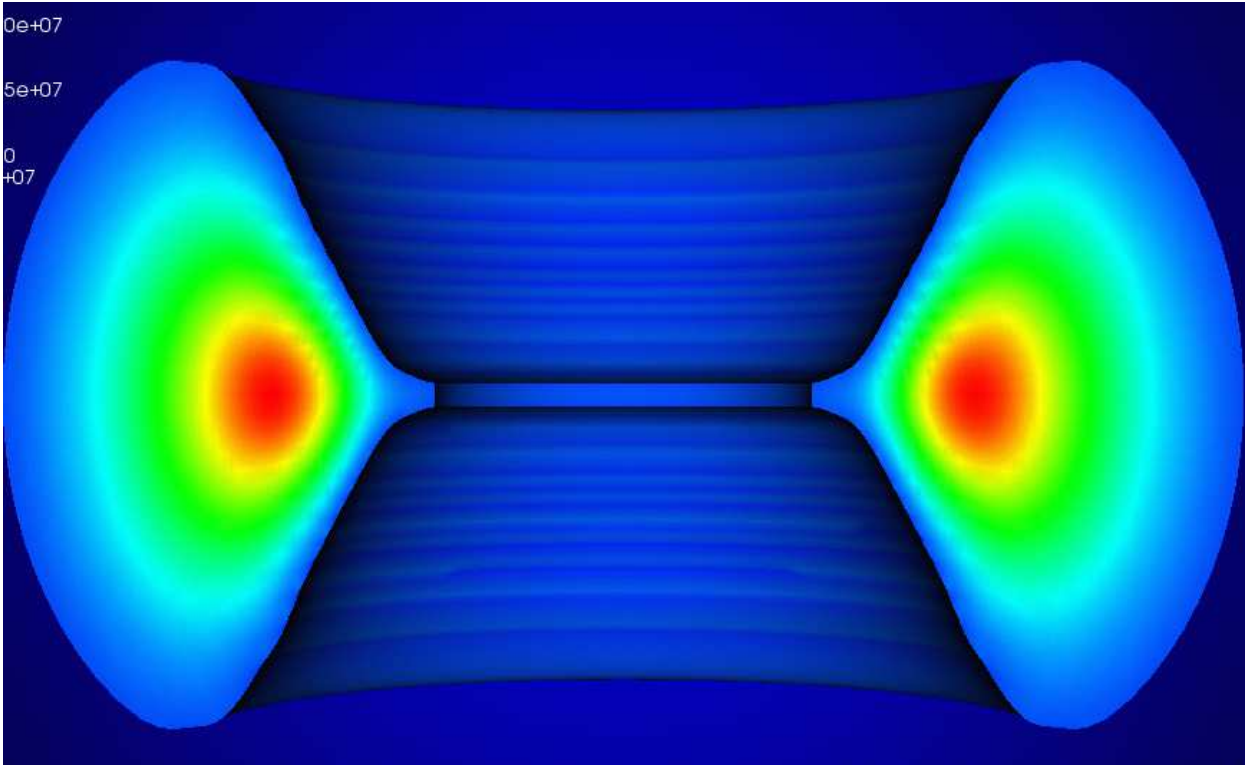}
\caption{Model t90.
({\it a}) Linear density plot at $t=490M\approx10\tau_{orb}$.
For rendering purposes, the surface of the torus is set at
$\rho_{surf}\approx0.1\rho_{max}$ and half the torus is cut away.
To give a clearer view of the central
region, the image is zoomed in to a small fraction
of the total computational volume.  The image is plotted in logarithmic
coordinate space, where $x=\eta\sin\vartheta\cos\varphi$,
$y=\eta\sin\vartheta\sin\varphi$, and $z=\eta\cos\vartheta$.
The torus is only initialized in the region $r\ge r_{cusp}$.
The black hole (not pictured) lies at the center of the image.
({\it b}) Mass accretion rate $\dot{M}$ at the black hole horizon
in units of the Eddington mass accretion rate
$\dot{M}_{Edd}$ (assuming $M_{BH}=1 M_\odot$ and $\kappa=1.5\times 10^{20}$
in cgs units).
({\it c}) PPI mode growth.}
\label{fig:modelt90}
\end{figure}

This is the only model without an initial tilt between the symmetry
plane of the black hole and the midplane of the torus.  The black
hole has a spin $a/M=0.9$ and radius $r_{BH}/r_G=1.44$; the torus
has a specific angular momentum $l/M=2.6088$ and a density center at
$r_{center}/r_G=3.45$. The cusp in the potential is located at
$r_{cusp}/r_G=1.76$, $\theta=\pi/2$ with $\Phi_{cusp}=-0.04$. We
choose an inner surface potential $\Phi_{in}=-0.01$ giving an energy
gap $\Delta \Phi = 0.03$.  The outer boundary is set at
$r_{max}/r_G=80$ and the simulation is evolved until $t=490
M\approx10\tau_{orb}$. This untilted simulation is included
primarily to illustrate that our base model is not susceptible to
the Papaloizou-Pringle instability, owing to the relatively large,
positive energy gap at the cusp leading to a moderately high
accretion rate that suppresses the growth of PPI \citep{haw91}. In
Figure \ref{fig:modelt90} we show ({\it a}) the final distribution
of torus gas density, ({\it b}) the time evolution of the mass
accretion rate at the black hole horizon, and ({\it c}) the growth
rates for the $m=1$ and $m=2$ PPI modes.

\begin{figure}
\figurenum{2b}
\plotone{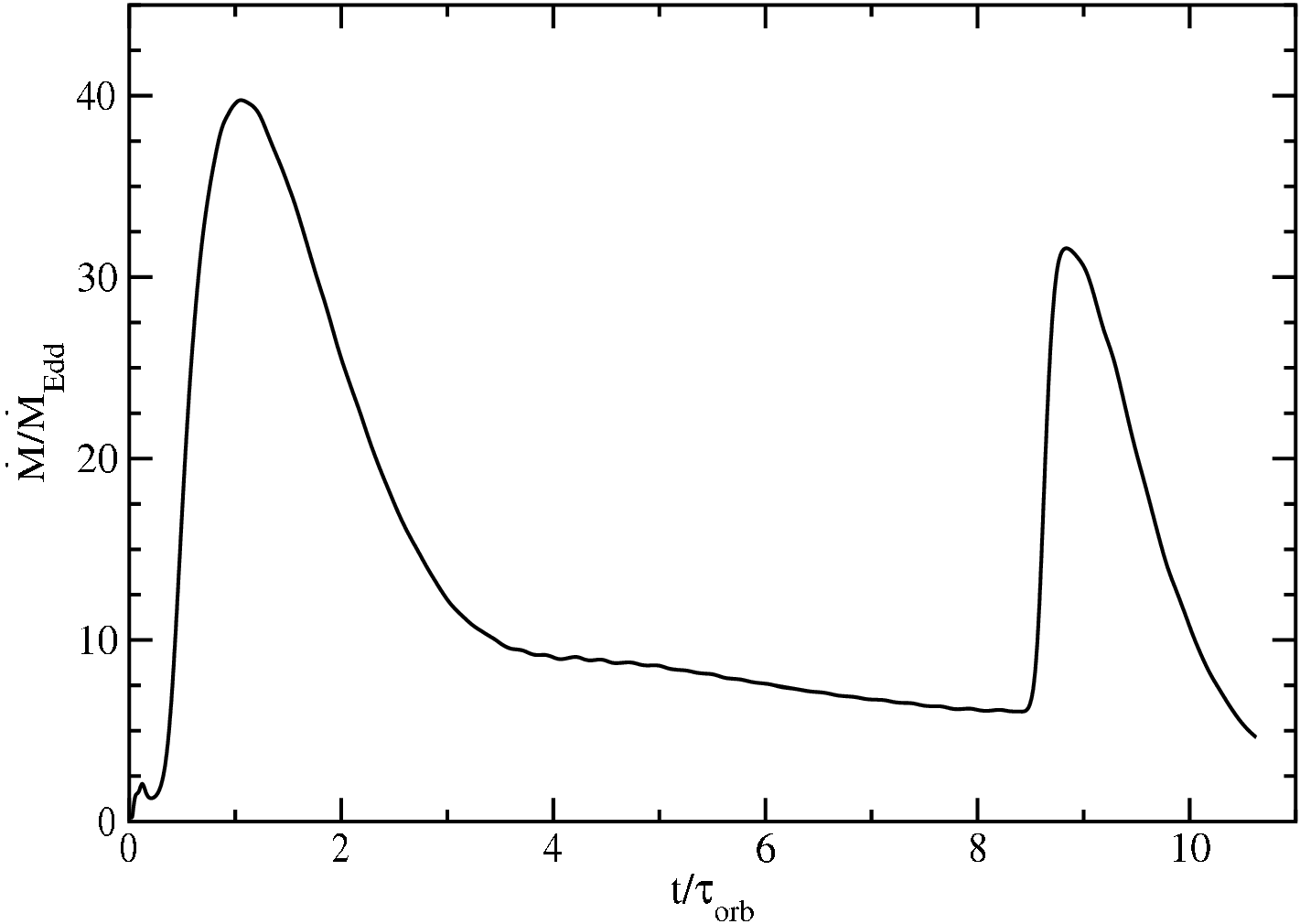}
\caption{}
\end{figure}

\begin{figure}
\figurenum{2c}
\plotone{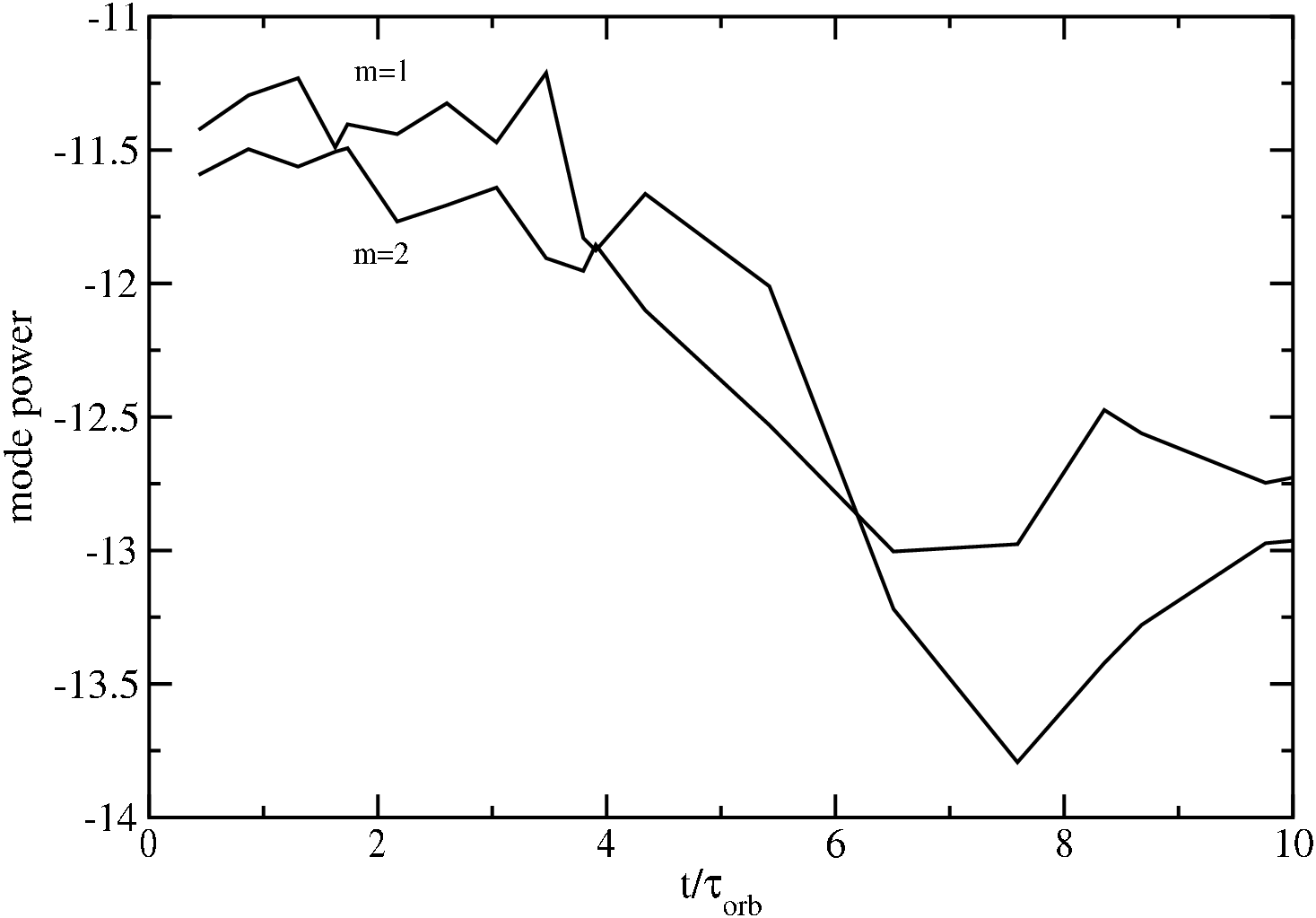}
\caption{}
\end{figure}

The mass accretion rate (Figure \ref{fig:modelt90}{\it b})
starts at a small value fixed by the background
Bondi solution.  Gas from the disk begins its infall from
$r>r_{cusp}$ and so experiences a short delay before reaching
the black hole horizon.  After a short transition
time ($\lesssim 3 t_{orb}$), $\dot{M}$ asymptotes toward a
stationary value.  However, the initial adjustment of the disk generates an
outgoing wave within the disk
that is reflected at the outer disk edge.  The arrival of this wave
back at the inner edge of the disk
causes the ``bump'' in the accretion rate at
late time.  A similar behavior is seen in many of our models.
Although this feature shows up very clearly in the accretion rate, it
does not appear to have a significant effect on any of the other
disk properties we track.  This is consistent with our findings
throughout this study that the tilt and twist of these disks are
not strongly responsive to the mass accretion rate.
Furthermore, we note from our longer runs below (see Model t915L)
that this feature does not recur at later times; after the wave
reaches the inner edge of the disk, it passes
through the cusp and black-hole horizon and is not reflected.
It, therefore, has no further impact on the problem.
Based upon this behavior, it appears that this feature is either not related
to the axisymmetric oscillation modes studied in \citet{zan03} or
that the large value of $\Delta \Phi$ (energy gap) prevents the
oscillation mode from being sustained.

Although the PPI
is seeded with small ($\lesssim1$\%) random fluctuations in the initial
mass and internal energy densities,
its growth is suppressed and
the torus becomes essentially smooth and axisymmetric.
This is consistent with the findings of \citet{dev02} that the
PPI is suppressed whenever the accretion rate is high as in our models.
In fact, this motivated our choice of parameters for these models, as
we wanted to avoid the complication of PPI in this study.

\subsection{Models t915XL, t915L, \& t915H}
\begin{figure}
\figurenum{3a}
\plotone{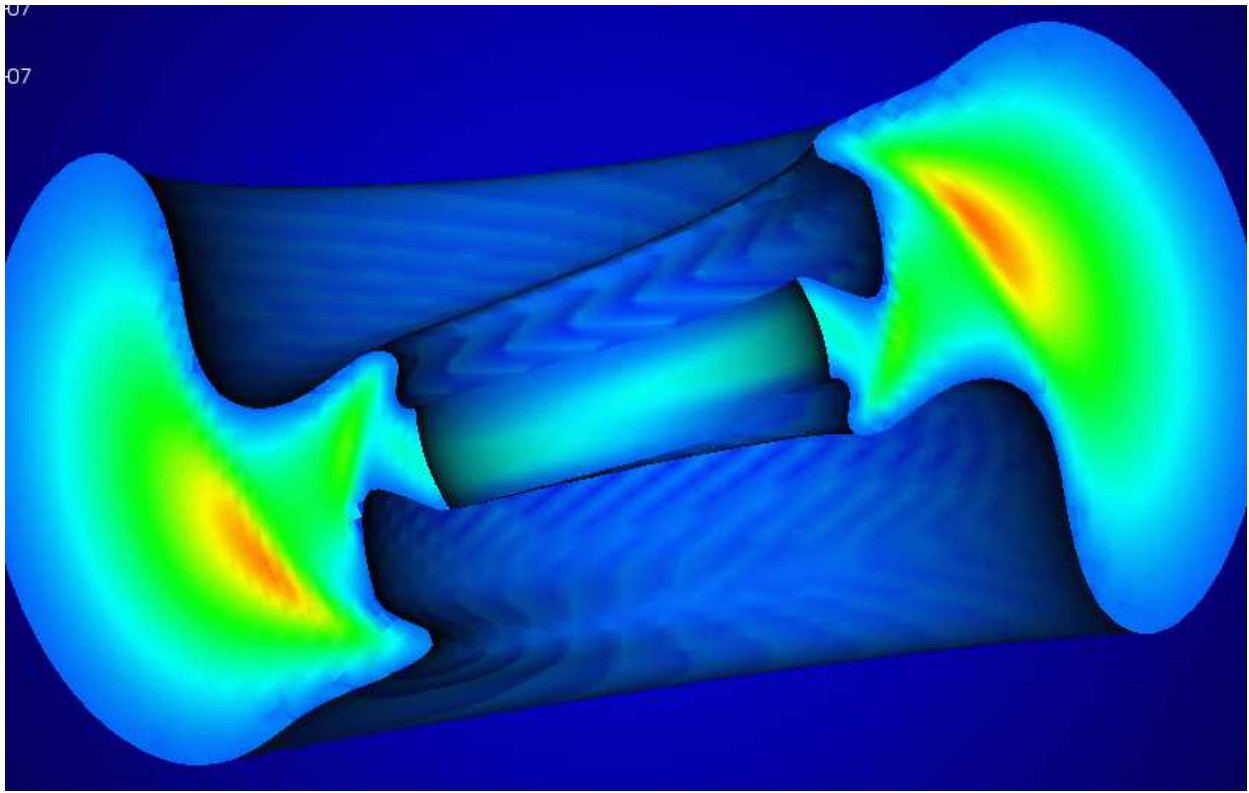}
\caption{Model t915L \& t915H.
({\it a}) Linear density profile (similar to Figure \ref{fig:modelt90}{\it a})
for Model t915H at $t=490M\approx10\tau_{orb}$.  The black hole spin axis is
tilted $15^\circ$ to the left of the $z$-axis.
({\it b}) Cumulative
precession angle $\gamma$ and ({\it c}) tilt $\beta$ in radians as functions
of radius.  Initially, $\gamma=0$ and
$\beta=15^\circ\approx0.26$ throughout the disks.  ({\it d}) Mass accretion
rate in units of the Eddington mass accretion rate.  ({\it e}) Transition
radius ({\it symbols}) as a function of time along with the Lense-Thirring
precession timescale $\tau_{LT}$ ({\it solid}), azimuthal sound crossing time
$\tau_{cs}$ ({\it dashed}), and accretion timescale
$\tau_{acc}$ ({\it dot-dashed}).  The thin vertical line marks the radius
where $\tau_{LT}=\tau_{cs}$.}
\label{fig:modelt915}
\end{figure}

These models begin from exactly the same initial conditions as Model
t90 except that the black hole is tilted by an angle
$\beta_0=15^\circ$.  Thus, these disks are subject to Lense-Thirring
precession. Model t915XL uses a resolution of $24\times24\times24$
zones and Model t915L uses a resolution of $48\times24\times48$;
both are evolved until $t=1400 M\approx30\tau_{orb}$.  Model t915H
uses a resolution of $96\times64\times96$ and is evolved until
$t=490 M\approx10\tau_{orb}$. Figure \ref{fig:modelt915} shows ({\it
a}) the final distribution of torus gas density for Model t915H,
({\it b}) the cumulative precession $\gamma$ and ({\it c}) tilt
$\beta$ of the disk as a function of the logarithmic radial
coordinate $\eta$, ({\it d}) the mass accretion rate $\dot{M}$ at
the black hole horizon as a function of time, and ({\it e}) the time
evolution of the transition radius $r_T$ defined as the radius where
the twist equals 1 radian ($\gamma=1$). It is clear from Figures
\ref{fig:modelt915}{\it a}, \ref{fig:modelt915}{\it b}, and
\ref{fig:modelt915}{\it c} that the disk is strongly twisted and
warped at the end of the simulation. Nevertheless, the disk reaches
a quasi-steady state before the simulations are stopped. This claim
is supported by the following observations: 1) in Figure
\ref{fig:modelt915}{\it b}, the twist profile changes only slightly
from $t=10\tau_{orb}$ to $t=30\tau_{orb}$; 2) in Figure
\ref{fig:modelt915}{\it d}, the mass accretion rate asymptotes
toward a constant value; and 3) in Figure \ref{fig:modelt915}{\it
e}, the transition radius also asymptotes toward a constant value.

\begin{figure}
\figurenum{3b}
\plotone{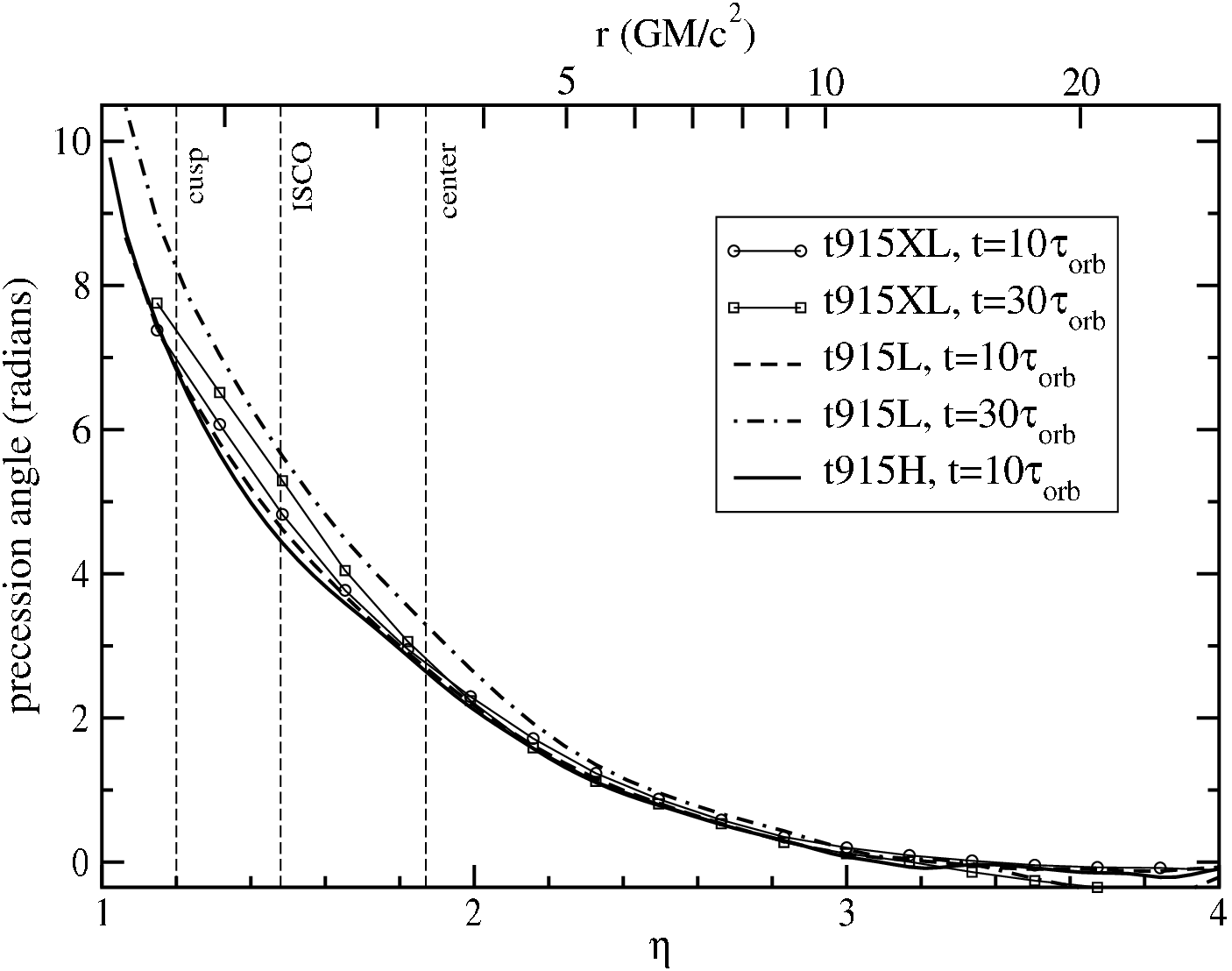}
\caption{}
\end{figure}

\begin{figure}
\figurenum{3c}
\plotone{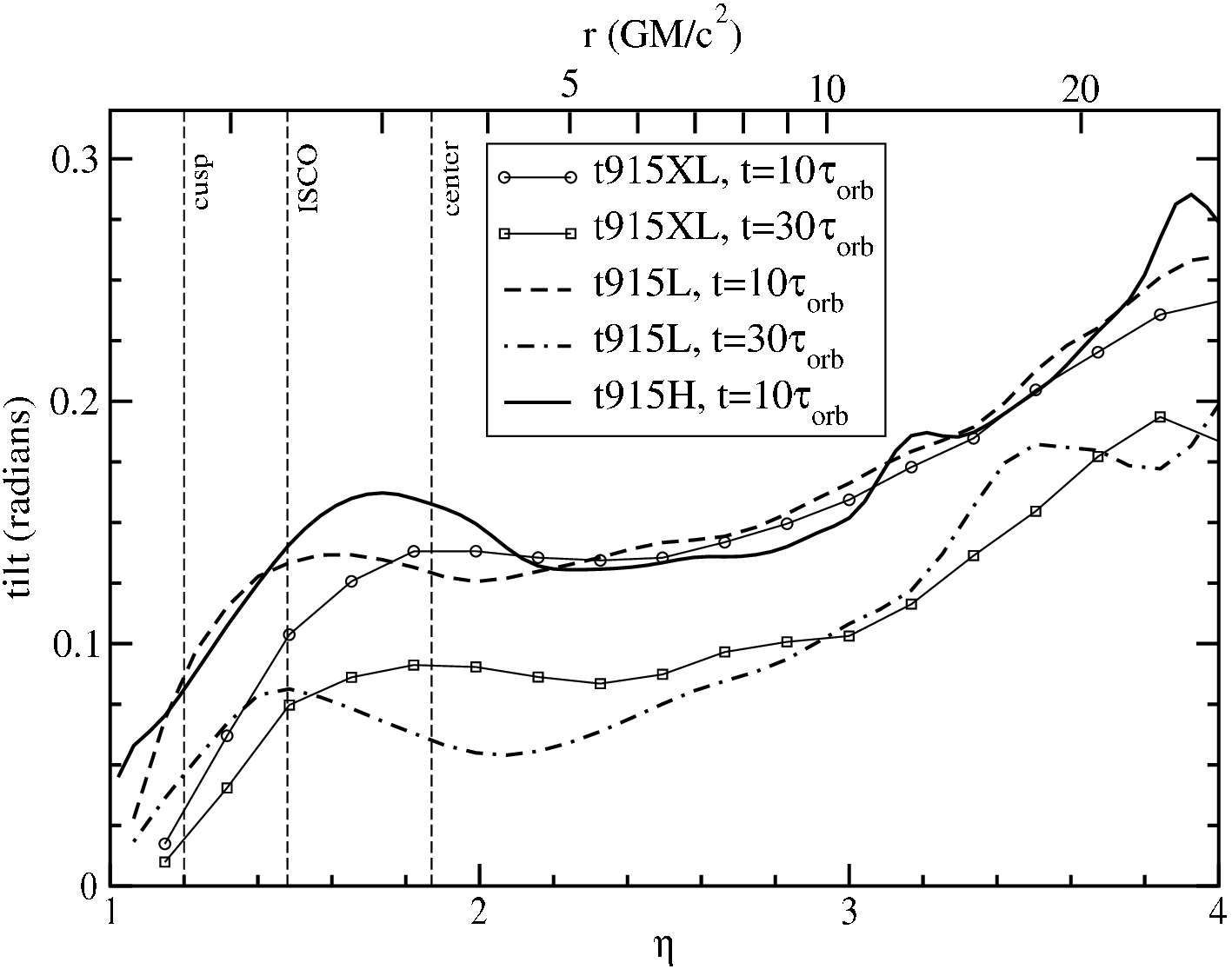}
\caption{}
\end{figure}

\begin{figure}
\figurenum{3d}
\plotone{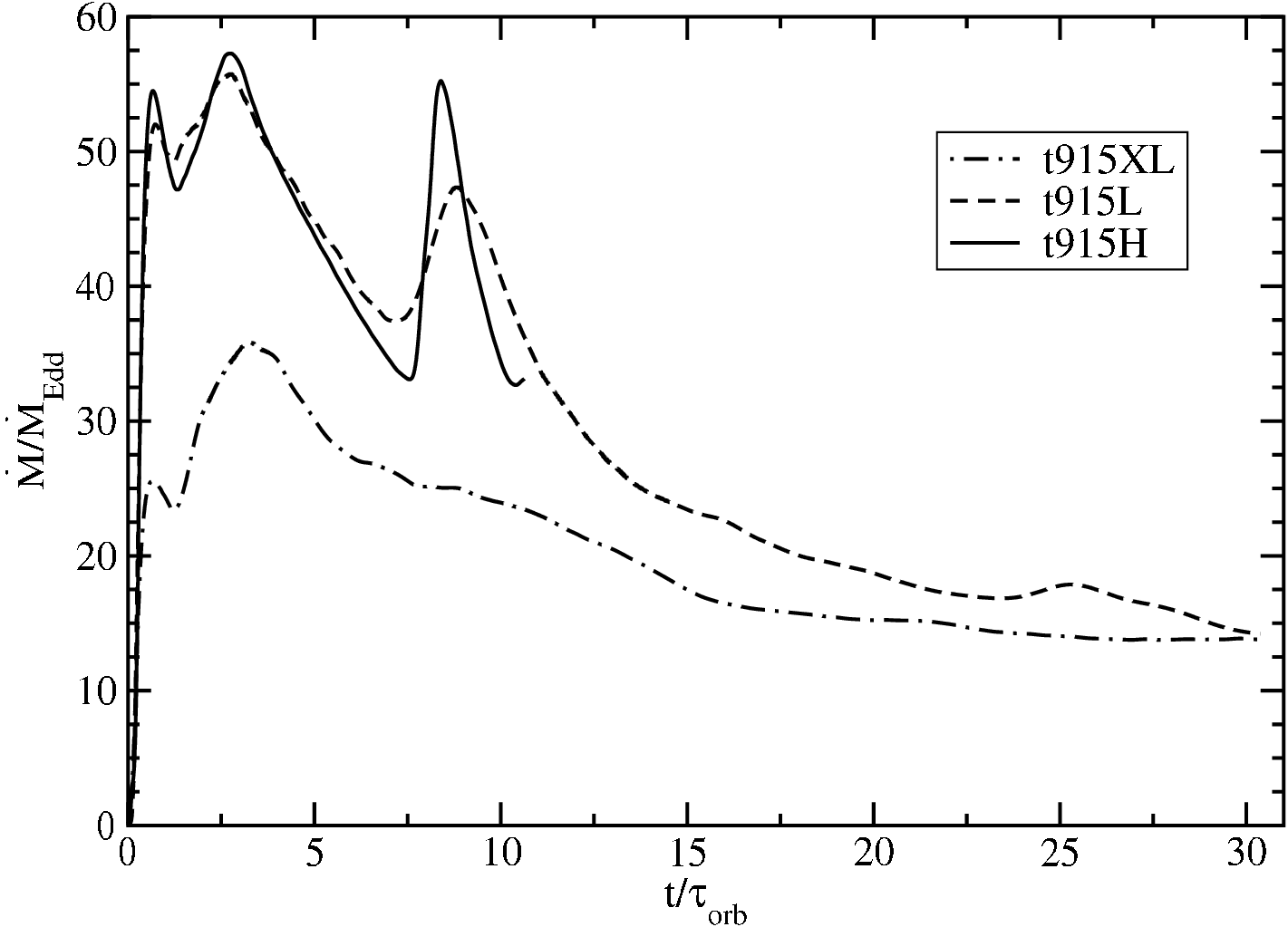}
\caption{}
\end{figure}

\begin{figure}
\figurenum{3e}
\plotone{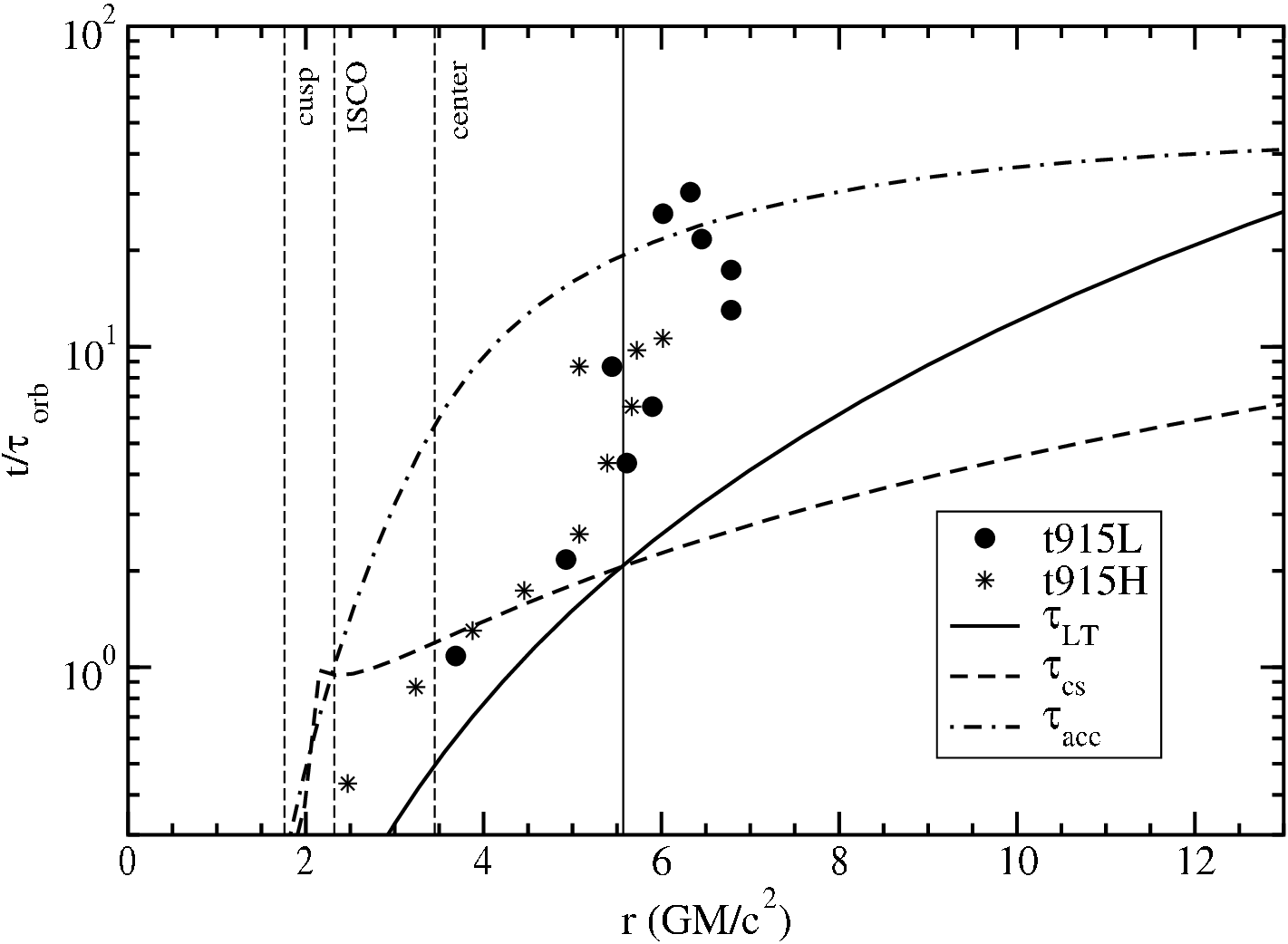}
\caption{}
\end{figure}

Although the transition radius in Figure \ref{fig:modelt915}{\it e}
starts out closely following the expected growth of Lense-Thirring
precession for ideal test particles ($r_T\propto t^{1/3}$), it
eventually approaches a limiting value of $r_T \lesssim 7 r_G$. This
is approximately coincident with the radius at which the local
azimuthal sound-crossing time $\tau_{cs}$ ({\it dashed line} in
Figure \ref{fig:modelt915}{\it e}) becomes shorter than the
Lense-Thirring precession timescale $\tau_{LT}$ ({\it thick solid
line}). This finding appears consistent with the expectations of
\citet{nel00} that warps in thick disks are diffused by wave modes.
As the disk is twisted up by Lense-Thirring precession, the twist is
diffused away by dynamical responses in the disk. We note that the
accretion timescale $\tau_{acc}$ ({\it dot-dashed line}) in these
disks is significantly longer than the Lense-Thirring or
sound-crossing timescales except in the outer regions 
($r>15 r_G$).  As such, it does not appear to play a role in
limiting the precession. Nevertheless, we point out that this tilted
disk is subject to a higher accretion rate than the untilted model
(compare Figures \ref{fig:modelt915}{\it d} and
\ref{fig:modelt90}{\it b}).  This is expected since more of the
tilted torus begins the simulation out of equilibrium.

A curious feature seen in Figure \ref{fig:modelt915}{\it c} and 
evident in nearly all of our models is the tendency
for these disks to align toward the symmetry plane of the black hole
despite the lack of viscous angular momentum transport.  This
alignment seems to be facilitated by the preferential accretion of
highly tilted disk material.  Notice in Figure
\ref{fig:modelt915}{\it a} that two opposing accretion streams start
from latitudes well above and below the black-hole symmetry plane.  The 
start of the lower stream is visible near the bottom of the disk.  The 
upper stream actually begins in the portion of the disk that is cut 
away, but this stream is nevertheless prominently visible cutting across 
the upper half of the disk.  
These streams begin near the density center of the disk and are 
$180^\circ$ out of phase with one another. The
material in these streams has a very large tilt relative to the black
hole and has a larger tilt, on average, than the material slightly further
out in the disk. This explains the prominent increase in tilt toward 
smaller radii seen near 
$r_{center}$ in Figure \ref{fig:modelt915}{\it c}.  Closer to the 
black hole, these two accretion streams spiral around 
until they cross near the symmetry plane of the black hole. 
As they cross the collisional dissipation of angular momentum produces a 
small ring of gas orbiting very near the
symmetry plane of the black hole. This ring has a very small
tilt, on average, and is responsible for 
the small value of $\beta$ seen near the black
hole horizon in Figure \ref{fig:modelt915}{\it c}.

The preferential accretion 
into the black hole of material with a large average tilt 
removes much of the
misaligned angular momentum from the disk, thus
allowing the system to gradually align.  For the low-mass disks 
considered here, it is primarily the disk orientation that changes.  
However, for sufficiently massive disks, such accretion 
could gradually change the orientation of the black hole spin axis.

We note from all of the figures that there is good agreement between
the runs at different resolutions.  This is both reassuring and
convenient as it allows us to use the course grid to follow the
evolution of these models for longer periods at less computational
expense.  We also note that the results appear to be converging
since the results of our intermediate resolution model (t915L) 
agree more closely with those of the high resolution model 
(t915H) than the low resolution one (t915XL).  Nevertheless, we acknowledge that there
may be features which are still not resolved even with our finest
grid.

\subsection{Model t915H2}
The initialization of Model t915H2 differs from
Model t915H above primarily in that
Model t915H2 has a larger initial energy gap at the cusp
($\Delta \Phi = 0.04$).  The larger energy gap implies a higher mass
accretion rate and a larger initial torus.
The torus (and the grid) for Model t915H2 are about three
times larger than in Model t915H ($r_{max}/r_G=240$).  This model is
resolved with the same high-resolution grid as Model t915H
($96\times64\times96$ zones), so the effective resolution in the
radial direction is reduced by about a factor of three.

\begin{figure}
\figurenum{4a}
\plotone{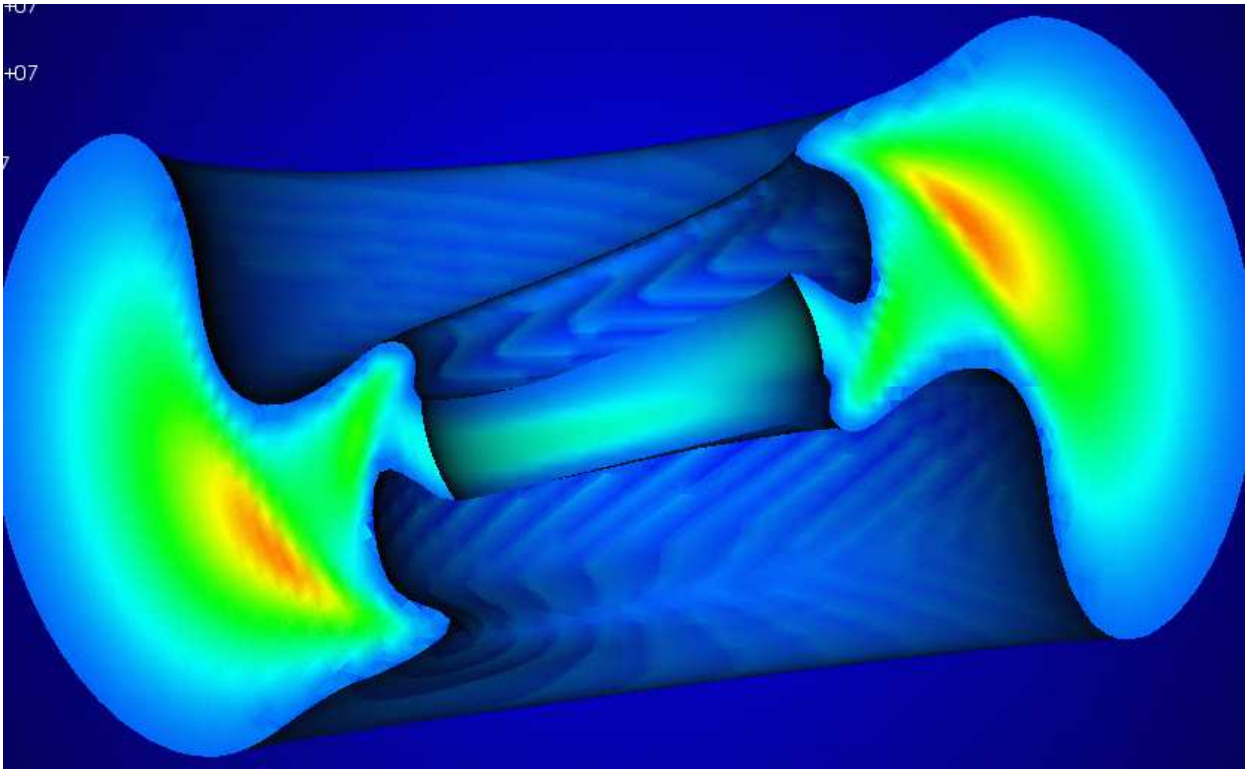}
\caption{Model t915H2.
({\it a}) Linear density profile (similar to Figure \ref{fig:modelt90}{\it a})
at $t=470M\approx10\tau_{orb}$.  The black hole spin axis is
tilted $15^\circ$ to the left of the $z$-axis.
({\it b}) Cumulative
precession angle $\gamma$ and ({\it c}) tilt $\beta$ in radians as functions
of radius.  Initially, $\gamma=0$ and
$\beta=15^\circ\approx0.26$ throughout the disks.  ({\it d}) Mass accretion
rate in units of the Eddington mass accretion rate.  ({\it e}) Transition
radius ({\it symbols}) as a function of time along with the Lense-Thirring
precession timescale $\tau_{LT}$, azimuthal sound crossing time
$\tau_{cs}$, and accretion timescale
$\tau_{acc}$.  The thin vertical line marks the radius
where $\tau_{LT}=\tau_{cs}$.}
\label{fig:modelt915H2}
\end{figure}

Despite the higher accretion rate of this model
(compare Figures \ref{fig:modelt915H2}{\it d} and
\ref{fig:modelt915}{\it d}), the measured precession $\gamma$
(Figure \ref{fig:modelt915H2}{\it b})
and tilt $\beta$ (Figure \ref{fig:modelt915H2}{\it c})
are very similar to those of Models t915L and t915H.
The final density profile is also very similar (compare
Figures \ref{fig:modelt915H2}{\it a} and
\ref{fig:modelt915}{\it a}).
This suggests that the
accretion rate is not a dominant factor controlling the evolution
of these models.  The size of the torus also does not appear to
play a significant role.

\begin{figure}
\figurenum{4b}
\plotone{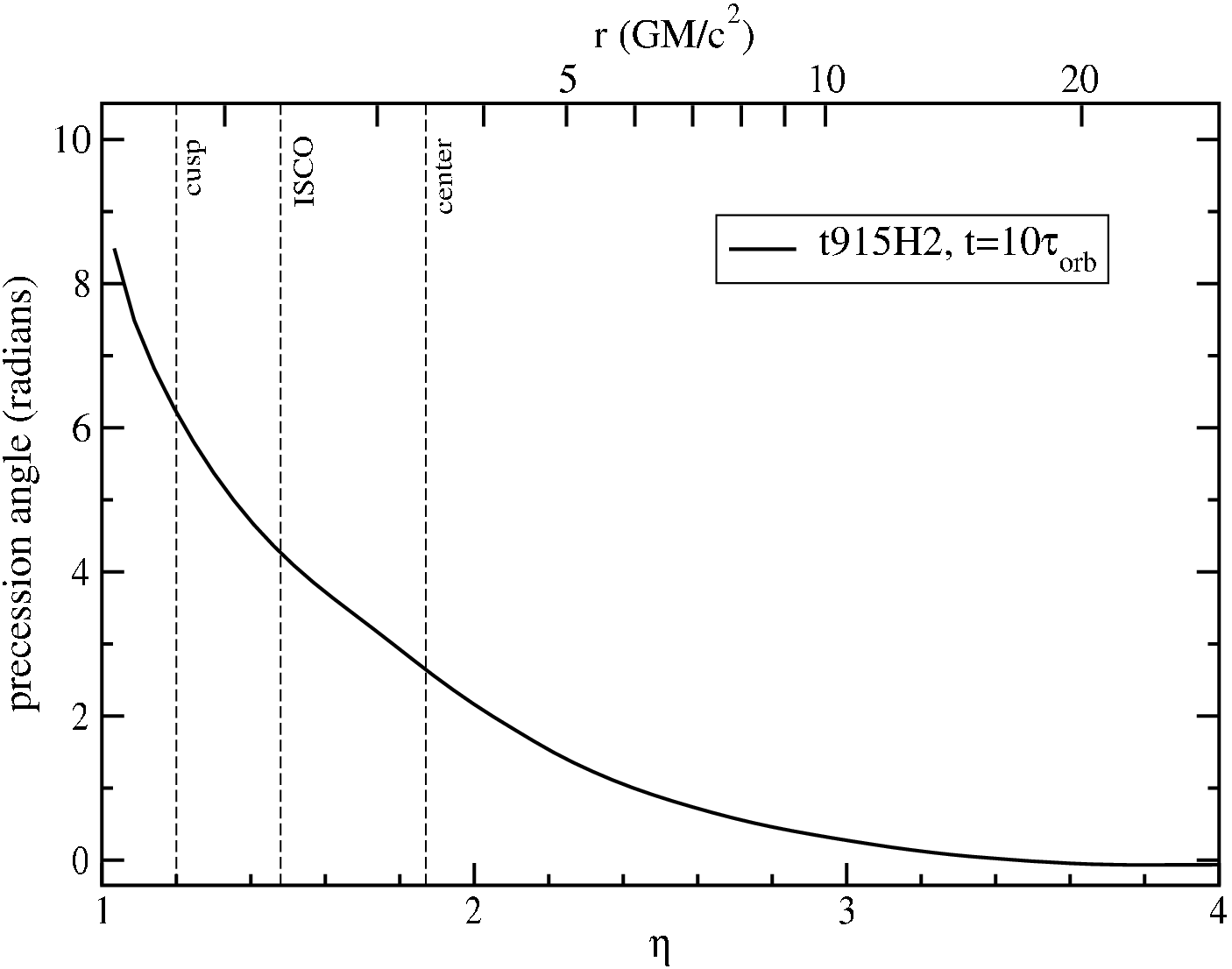}
\caption{}
\end{figure}

\begin{figure}
\figurenum{4c}
\plotone{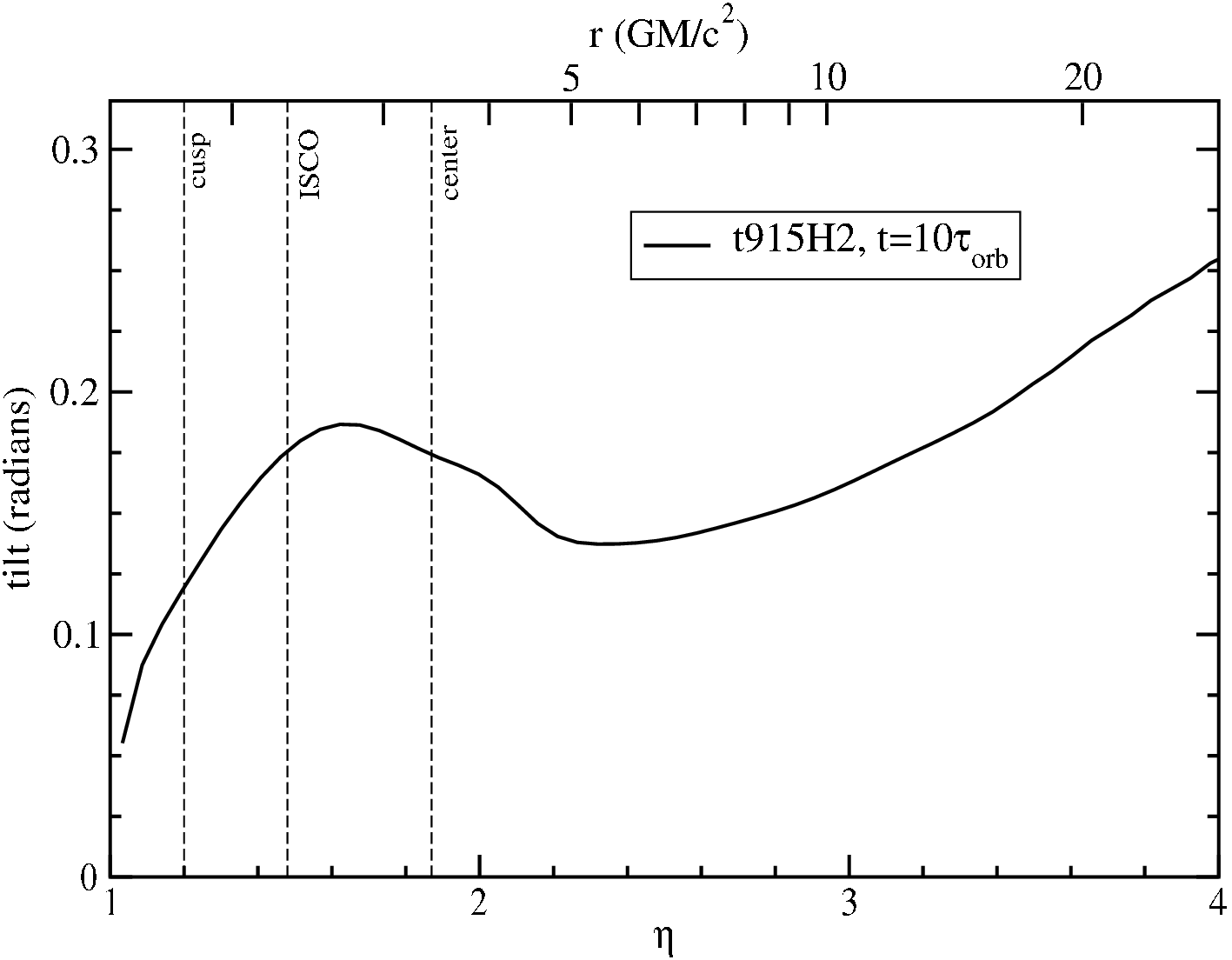}
\caption{}
\end{figure}

\begin{figure}
\figurenum{4d}
\plotone{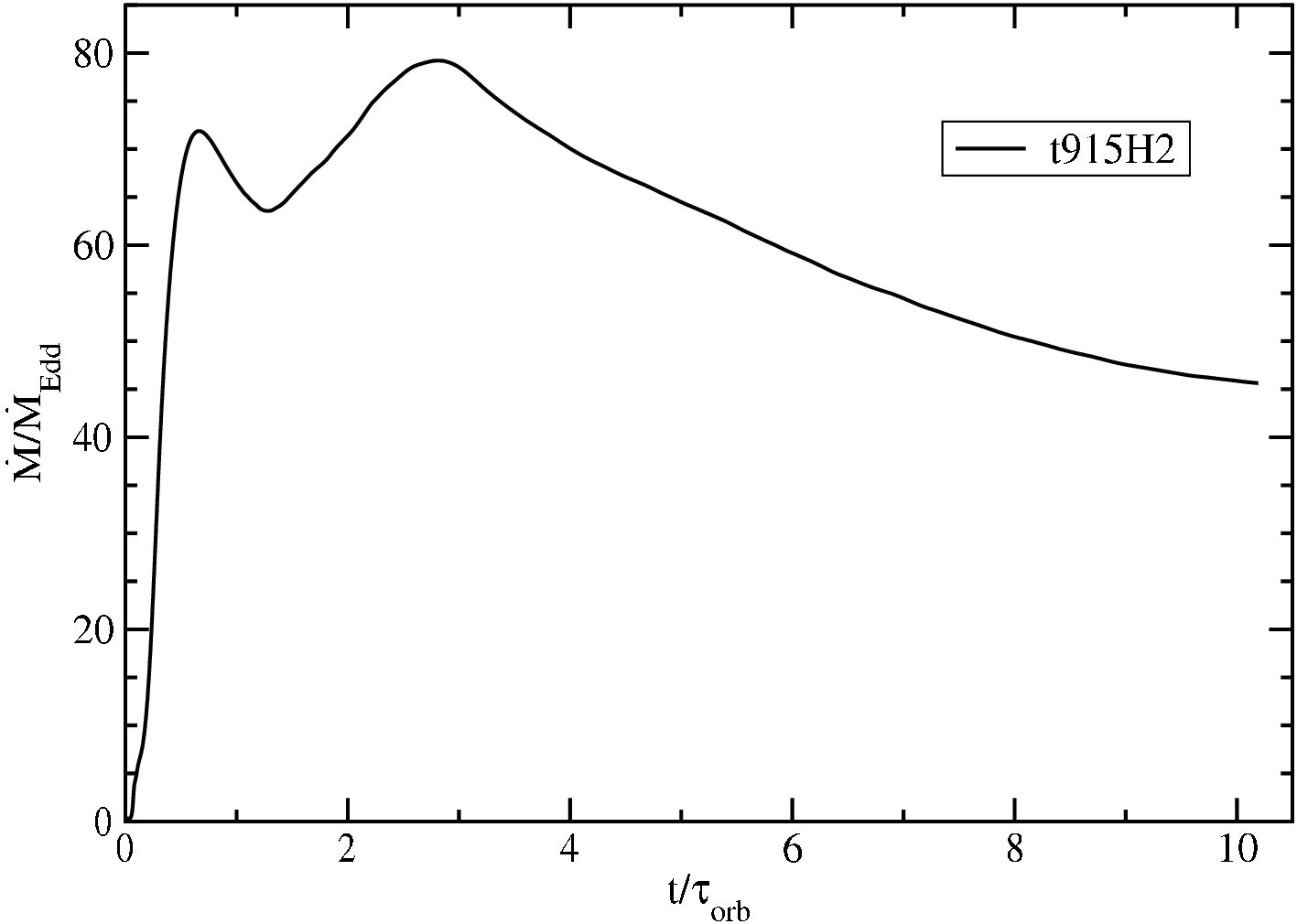}
\caption{}
\end{figure}

\begin{figure}
\figurenum{4e}
\plotone{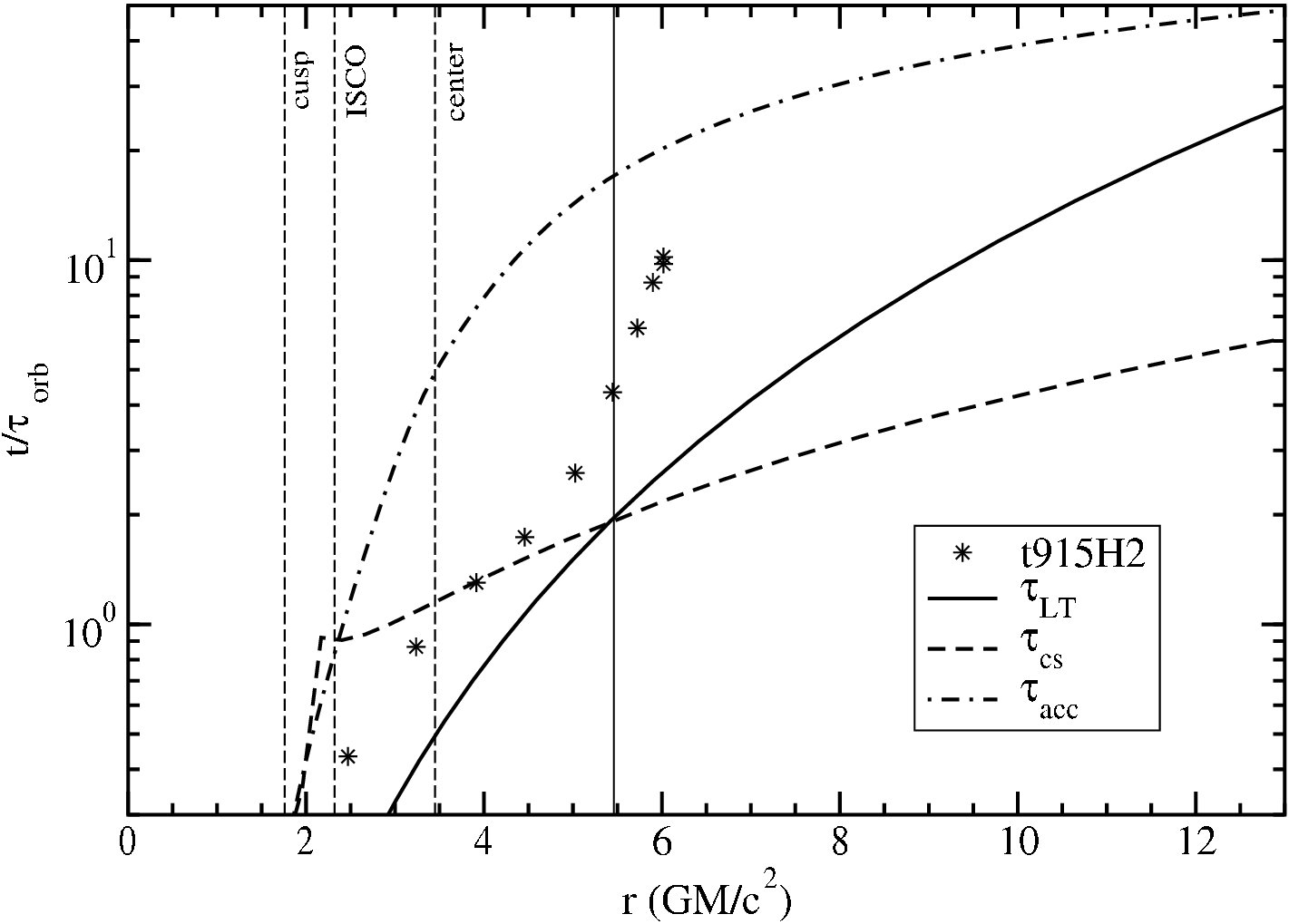}
\caption{}
\end{figure}

\subsection{Models t930L \& t930H}
These models also begin from exactly the same initial conditions as
Model t90 except that the black hole is tilted by an angle
$\beta_0=30^\circ$.  These models allow us to gauge how our results
depend on the initial tilt of the disk.
Model t930L uses a resolution of $48\times24\times48$ zones and is evolved
until $t=1400 M\approx30\tau_{orb}$;
Model t930H uses a resolution of $96\times64\times96$
and is evolved until $t=460 M\approx10\tau_{orb}$.
Again there is very good agreement between the two.

\begin{figure}
\figurenum{5a}
\plotone{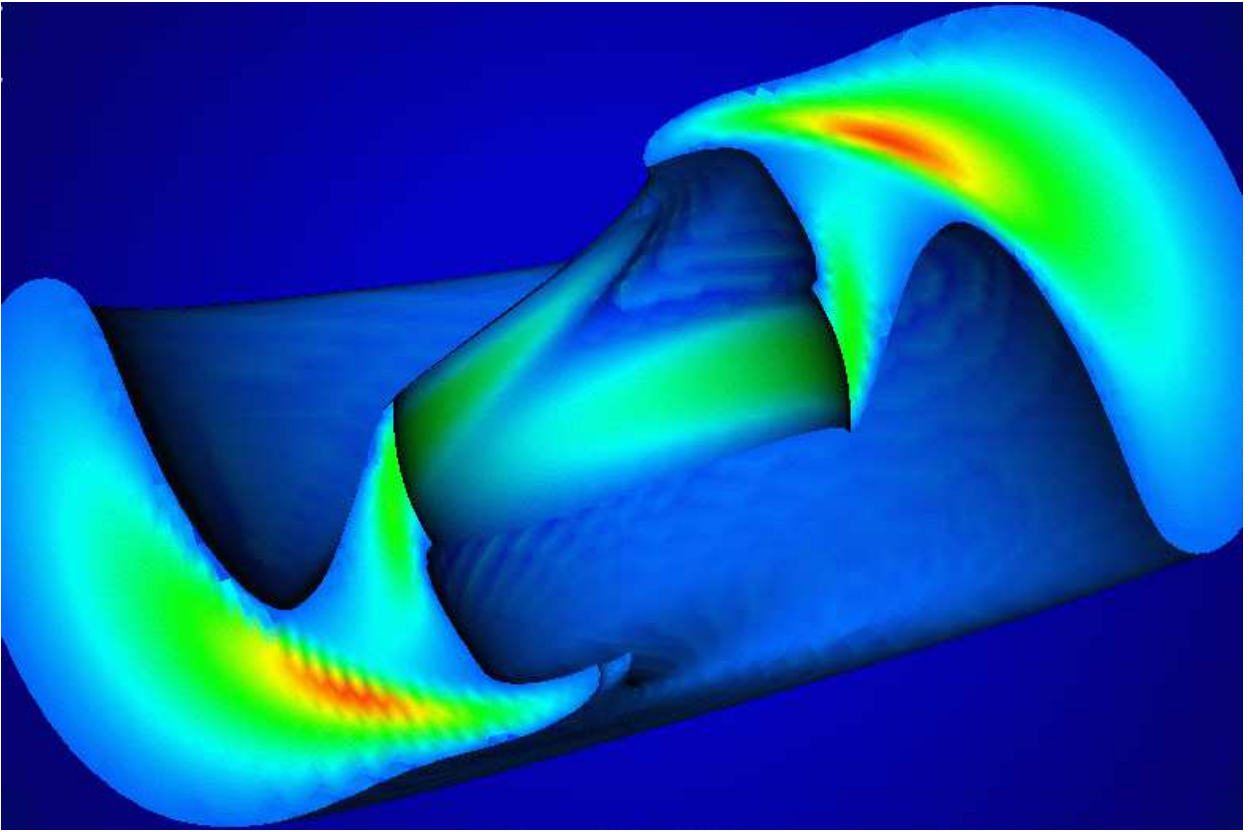}
\caption{Model t930L \& t930H.
({\it a}) Linear density profile (similar to Figure \ref{fig:modelt90}{\it a})
for Model t930H at $t=460M\approx10\tau_{orb}$.  The black hole spin axis is
tilted $30^\circ$ to the left of the $z$-axis.
({\it b}) Cumulative
precession angle $\gamma$ and ({\it c}) tilt $\beta$ in radians as functions
of radius.  Initially, $\gamma=0$ and
$\beta=30^\circ\approx0.52$ throughout the disks.  ({\it d}) Mass accretion
rate in units of the Eddington mass accretion rate.  ({\it e}) Transition
radius ({\it symbols}) as a function of time along with the Lense-Thirring
precession timescale $\tau_{LT}$, azimuthal sound crossing time
$\tau_{cs}$, and accretion timescale $\tau_{acc}$.
The thin vertical line marks the radius
where $\tau_{LT}=\tau_{cs}$.}
\label{fig:modelt930}
\end{figure}

These models have the highest accretion rates (Figure
\ref{fig:modelt930}{\it d}) of any considered.  Again, this is
expected since they begin with the disk furthest from equilibrium.
One consequence of the very high accretion rate is that the disk is
largely depleted of gas by the end of the simulation, particularly
for the longer evolution of Model t930L, which was carried out to
$t=1400 M\approx 30 \tau_{orb}$.  Nevertheless, we find very similar
results for the twist and tilt of these $\beta_0=30^\circ$ models
compared with the previous $\beta_0=15^\circ$ models (compare
Figures \ref{fig:modelt930}{\it c} and \ref{fig:modelt915}{\it c}).
We again see the preferential accretion of highly tilted gas through
two opposing streams starting near $r_{center}$.
Again these streams cross near the symmetry plane of the black hole, 
leaving a small, nearly aligned ring of gas close to the horizon. 
These results
further confirm that the mass accretion rate plays little, if any,
role in regulating the precession or tilt of these disks.

\begin{figure}
\figurenum{5b}
\plotone{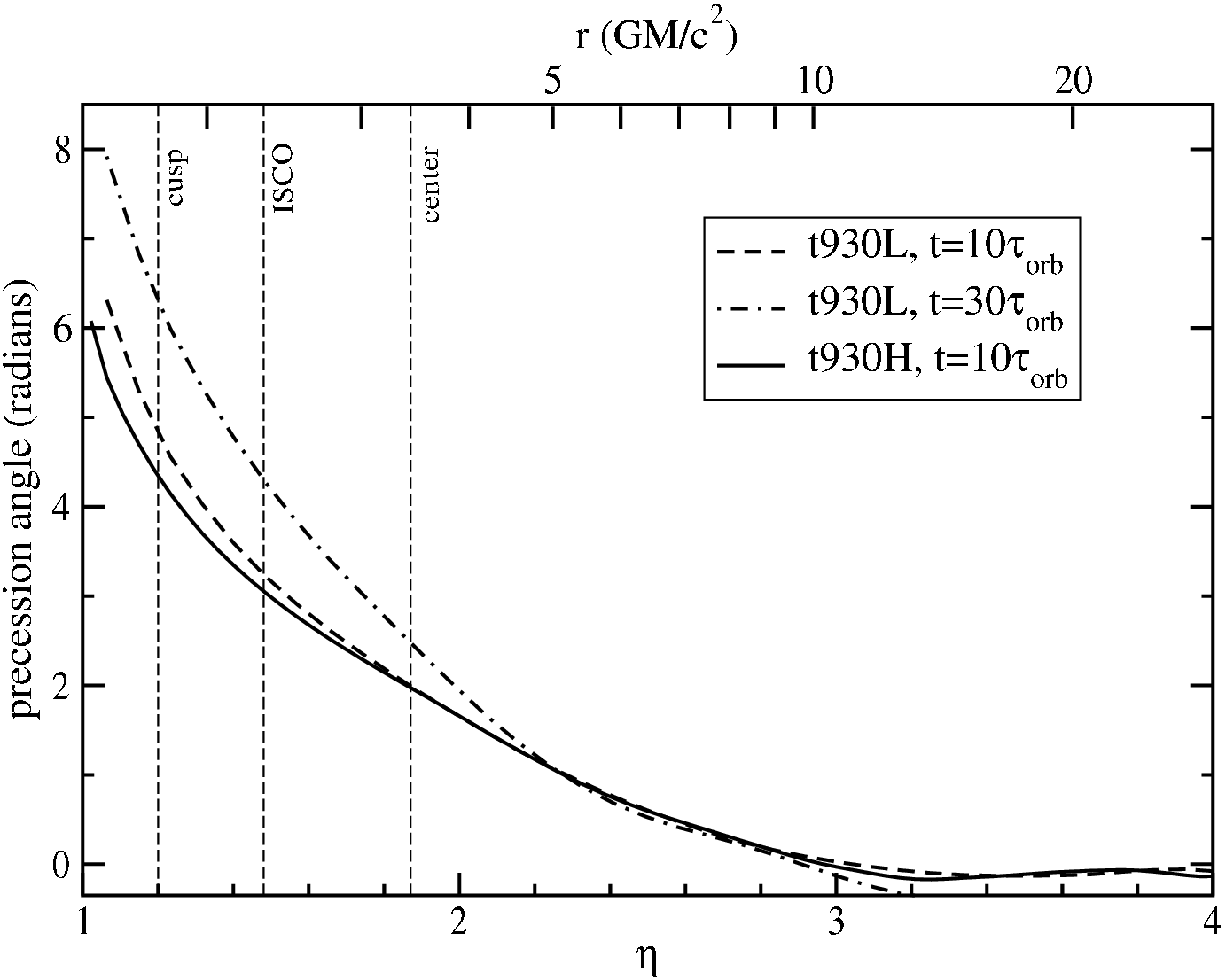}
\caption{}
\end{figure}

\begin{figure}
\figurenum{5c}
\plotone{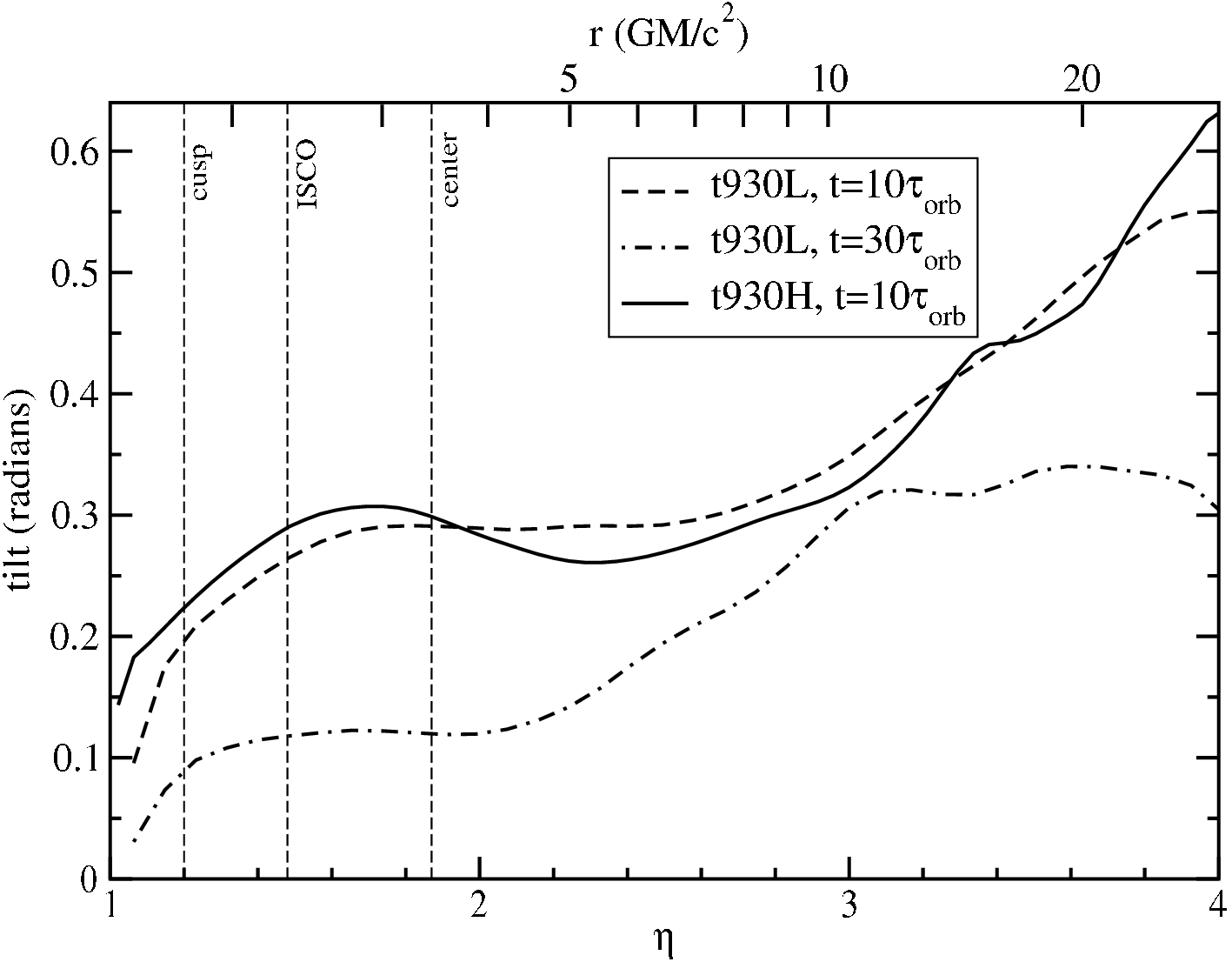}
\caption{}
\end{figure}

\begin{figure}
\figurenum{5d}
\plotone{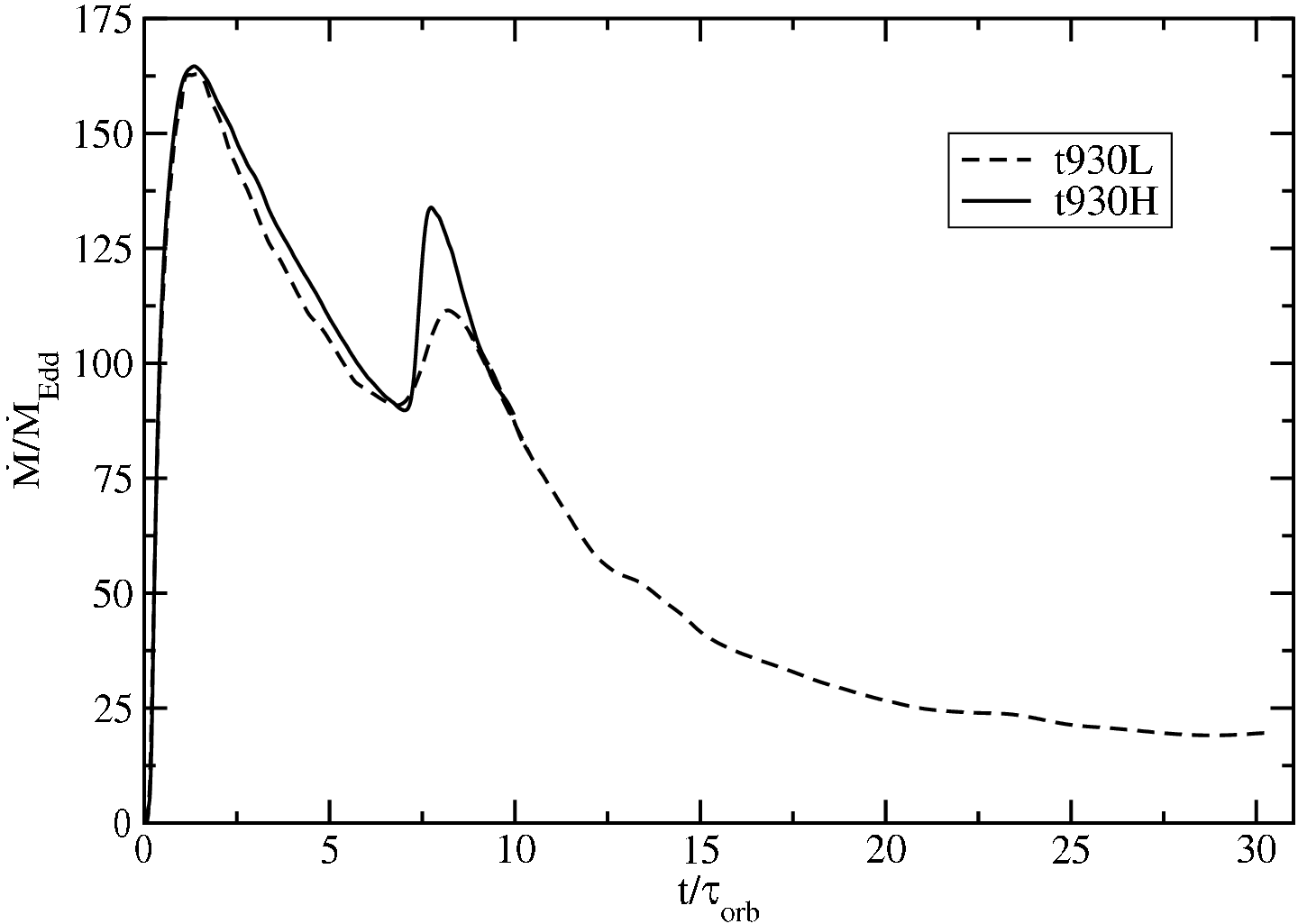}
\caption{}
\end{figure}

\begin{figure}
\figurenum{5e}
\plotone{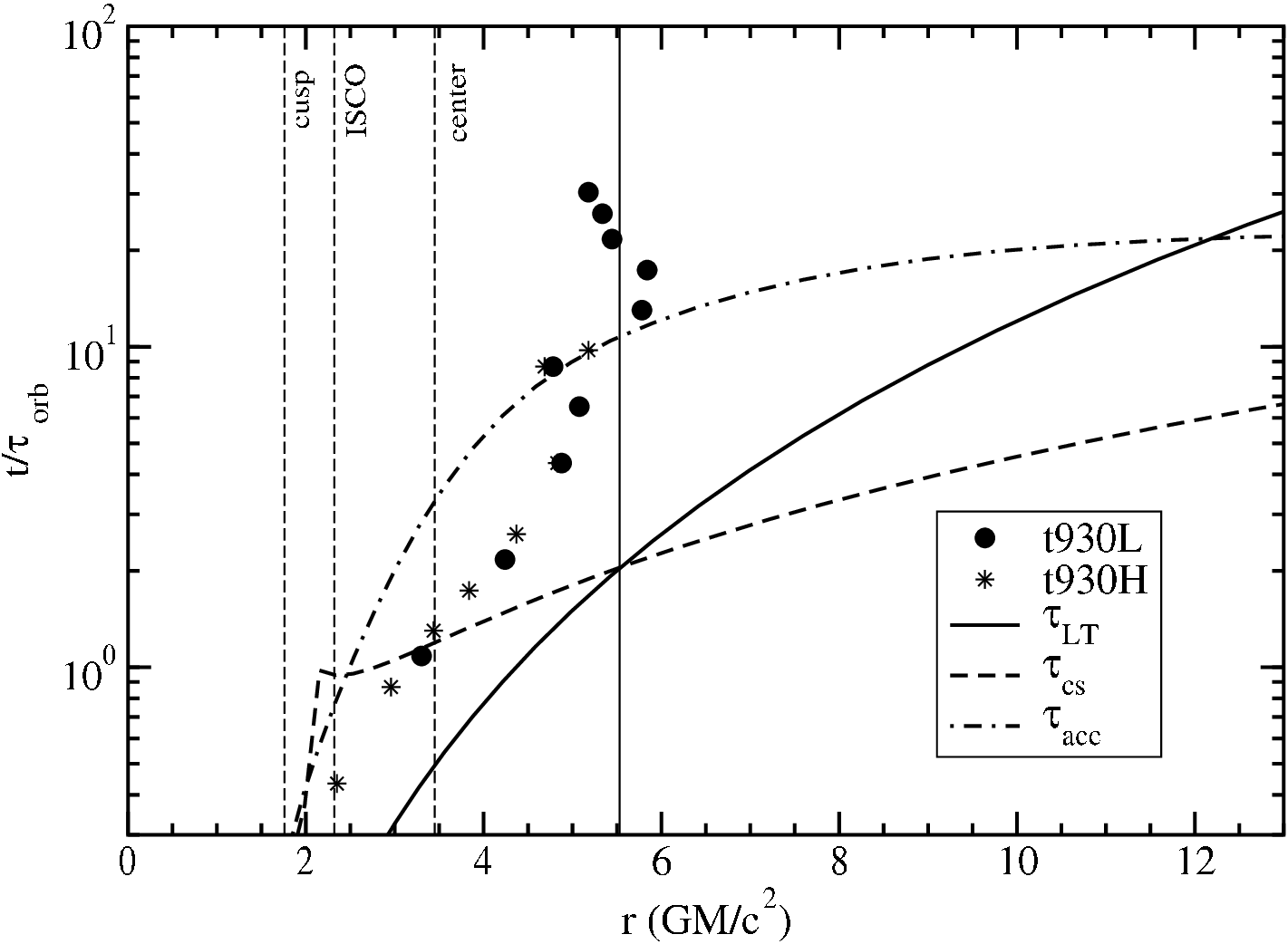}
\caption{}
\end{figure}

As in the previous models, the transition radius $r_T$ (Figure
\ref{fig:modelt930}{\it e}) asymptotes toward a radius $r_T\lesssim6
r_G$ consistent with the radius at which $\tau_{LT} \sim \tau_{cs}$.  The
early migration of $r_T$ is even similar to the previous models
(compare Figures \ref{fig:modelt930}{\it e} and
\ref{fig:modelt915}{\it e}).  This leaves the larger accretion rate
and tilt as practically the only differences between this and
previous models. Thus, it appears our results are only weakly
dependent on the initial tilt.

\subsection{Model t915$\Gamma$}
One of our general conclusions to this point is that differential 
Lense-Thirring precession in the disk is limited to radii 
for which $\tau_{LT}<\tau_{cs}$.  
However, this conclusion is based on a series of models that all 
share the same precession frequency and 
roughly the same sound speed in the disk.  It is useful, therefore, 
to consider a model where at least one of these timescales is 
changed. In Model t915$\Gamma$, we use the same black-hole spin 
($a/M=0.9$), 
disk parameters ($l/M=2.6088$ and $\Phi_{in}=-0.01$), and 
resolution ($48\times24\times48$) as Model t915L, except we change 
the equation of state of the gas such that 
$\Gamma=5/3$ and $\kappa=1.24\times10^5$ (in cgs units).  
Since the sound speed in these 
models scales as $(\Gamma-1)^{1/2}$, this change increases the 
sound speed in the disk by $\sqrt{2}$ and decreases the azimuthal 
sound-crossing time by the same factor.  

\begin{figure}
\figurenum{6a}
\plotone{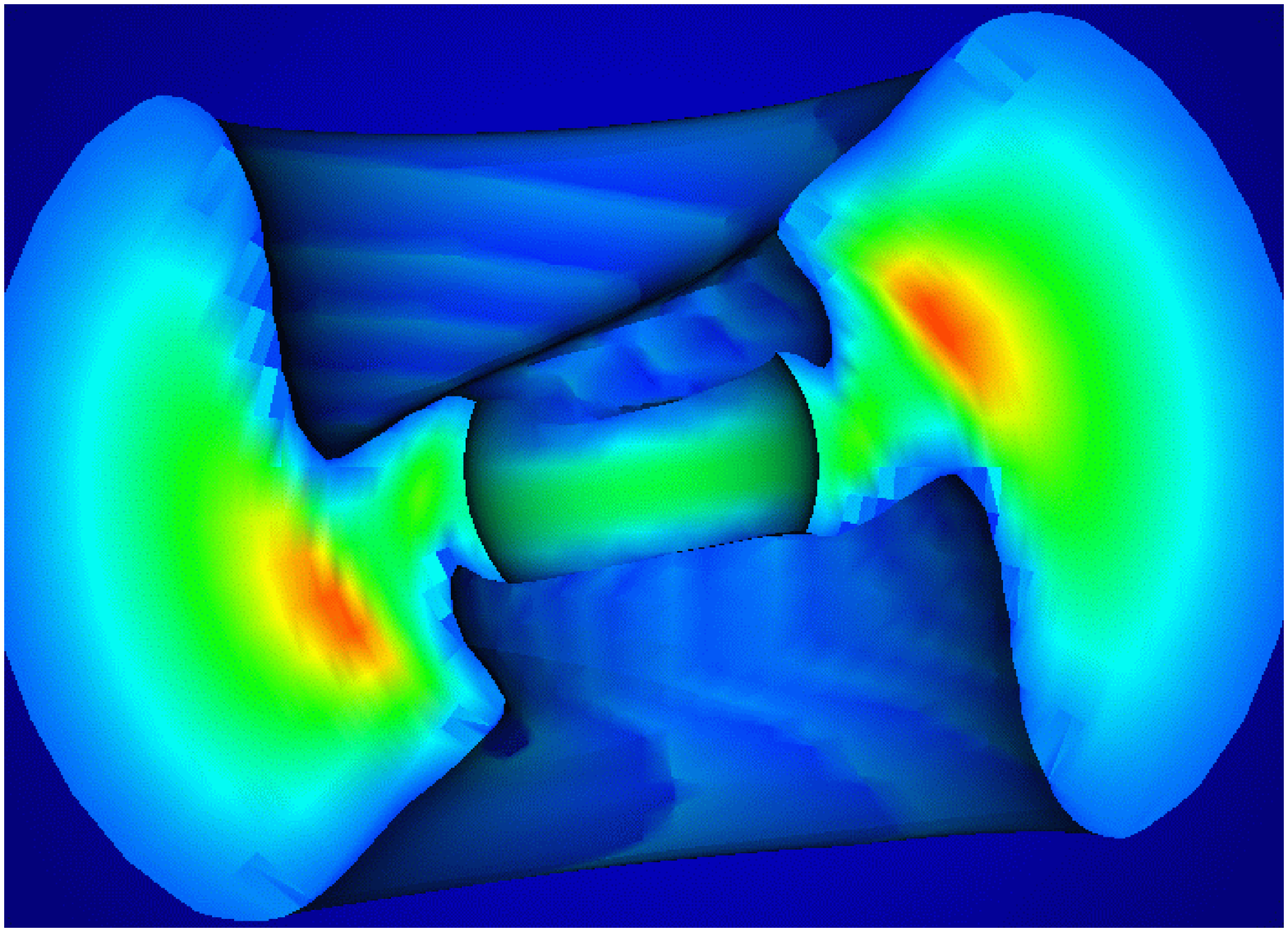}
\caption{Model t915$\Gamma$.
({\it a}) Linear density profile (similar to Figure \ref{fig:modelt90}{\it a})
at $t=1400M\approx30\tau_{orb}$.  The black hole spin axis is
tilted $15^\circ$ to the left of the $z$-axis.
({\it b}) Cumulative
precession angle $\gamma$ and ({\it c}) tilt $\beta$ in radians as functions
of radius.  Initially, $\gamma=0$ and
$\beta=15^\circ\approx0.26$ throughout the disks.  ({\it d}) Mass accretion
rate scaled to allow direct comparison with our other models.  
({\it e}) Transition
radius ({\it symbols}) as a function of time along with the Lense-Thirring
precession timescale $\tau_{LT}$, azimuthal sound crossing time
$\tau_{cs}$, and accretion timescale
$\tau_{acc}$.  The thin vertical line marks the radius
where $\tau_{LT}=\tau_{cs}$.}
\label{fig:modelt915G}
\end{figure}

All other things being equal, 
our first expectation is that the faster sound speed in this disk should 
make the transition radius 
asymptote at a smaller value than in the previous models.  Instead we 
see in Figure \ref{fig:modelt915G}{\it b} that the twist continues to 
grow with time throughout the disk.  This is also apparent 
in Figure \ref{fig:modelt915G}{\it e} where the 
transition radius $r_T$ 
gradually moves outward in the disk throughout the entire simulation 
and does not asymptote toward a constant value.  
By the end of the simulation it reaches a value more than double the radius 
where $\tau_{LT}=\tau_{cs}$.  
Although this result seems surprising in light of what we found in 
the previous models, when considered within the context of the results 
presented below, it is apparent that Model t915$\Gamma$ is a 
transition model.  In all of the previous models, the radius 
where $\tau_{LT}=\tau_{cs}$ lies outside of $r_{center}$; thus differential 
Lense-Thirring precession dominates the bulk of the disk material.  
In the models presented in the next two sections, the radius at which 
$\tau_{LT}=\tau_{cs}$ 
lies inside of $r_{center}$; those disks, therefore, have very efficient 
internal communication throughout the bulk of the disk and respond more 
uniformly to precession.  For the current model, $r_{center}$ and the radius 
at which $\tau_{LT}=\tau_{cs}$ differ by less than $25$\%; as a consequence, 
$\tau_{LT}\approx\tau_{cs}$ 
for the bulk of the disk material.  This balance prevents 
either differential precession or wave propagation from dominating 
the evolution.  As a result, in Figure \ref{fig:modelt915G}{\it e} 
we find an intermediate behavior between the 
vertical asymptote seen in strong differential precession models (Figures 
\ref{fig:modelt915}{\it e}, \ref{fig:modelt915H2}{\it e}, and 
\ref{fig:modelt930}{\it e}) and the nearly horizontal limit of solid body 
precession (Figure \ref{fig:modelt515}{\it e}).

\begin{figure}
\figurenum{6b}
\plotone{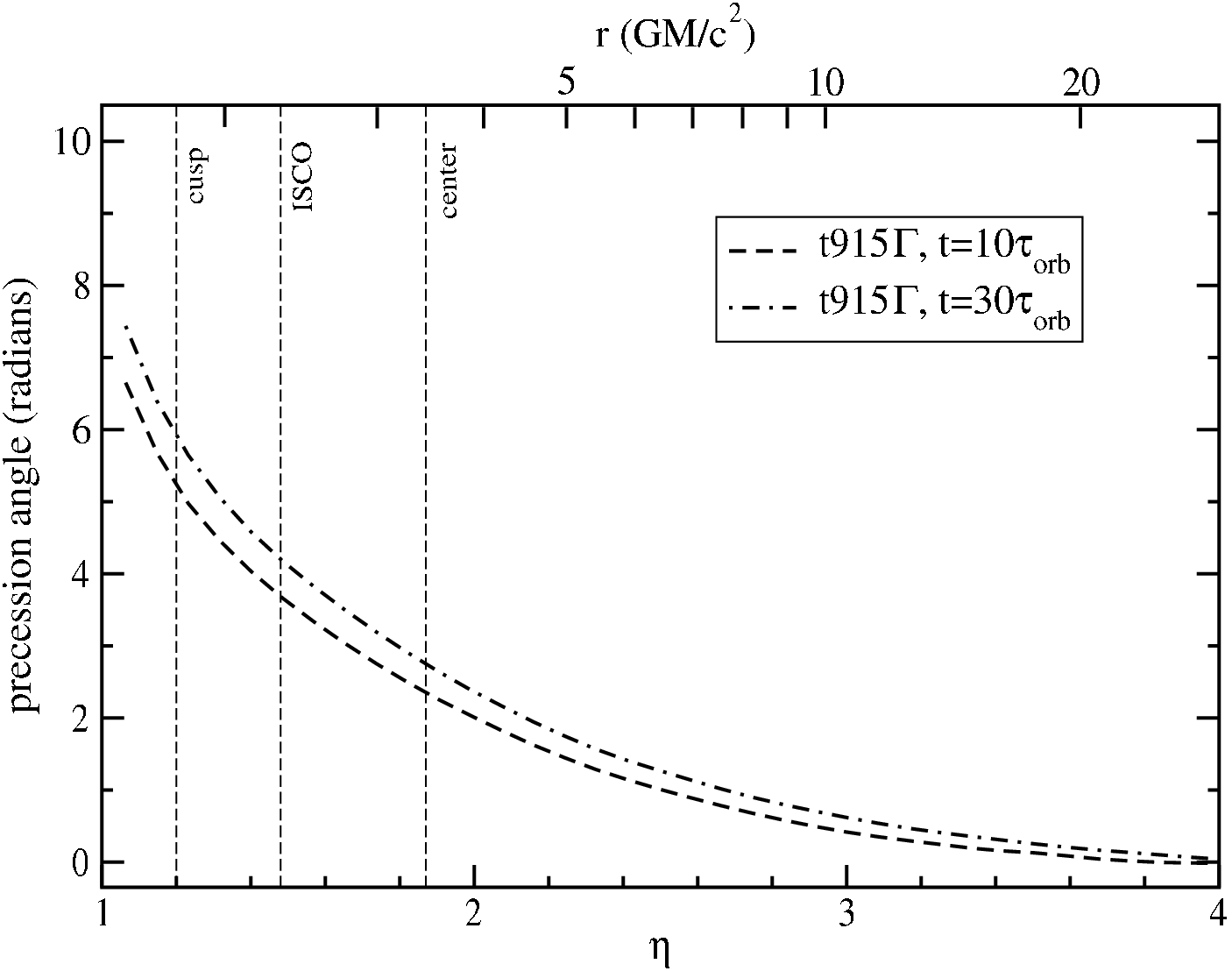}
\caption{}
\end{figure}

\begin{figure}
\figurenum{6c}
\plotone{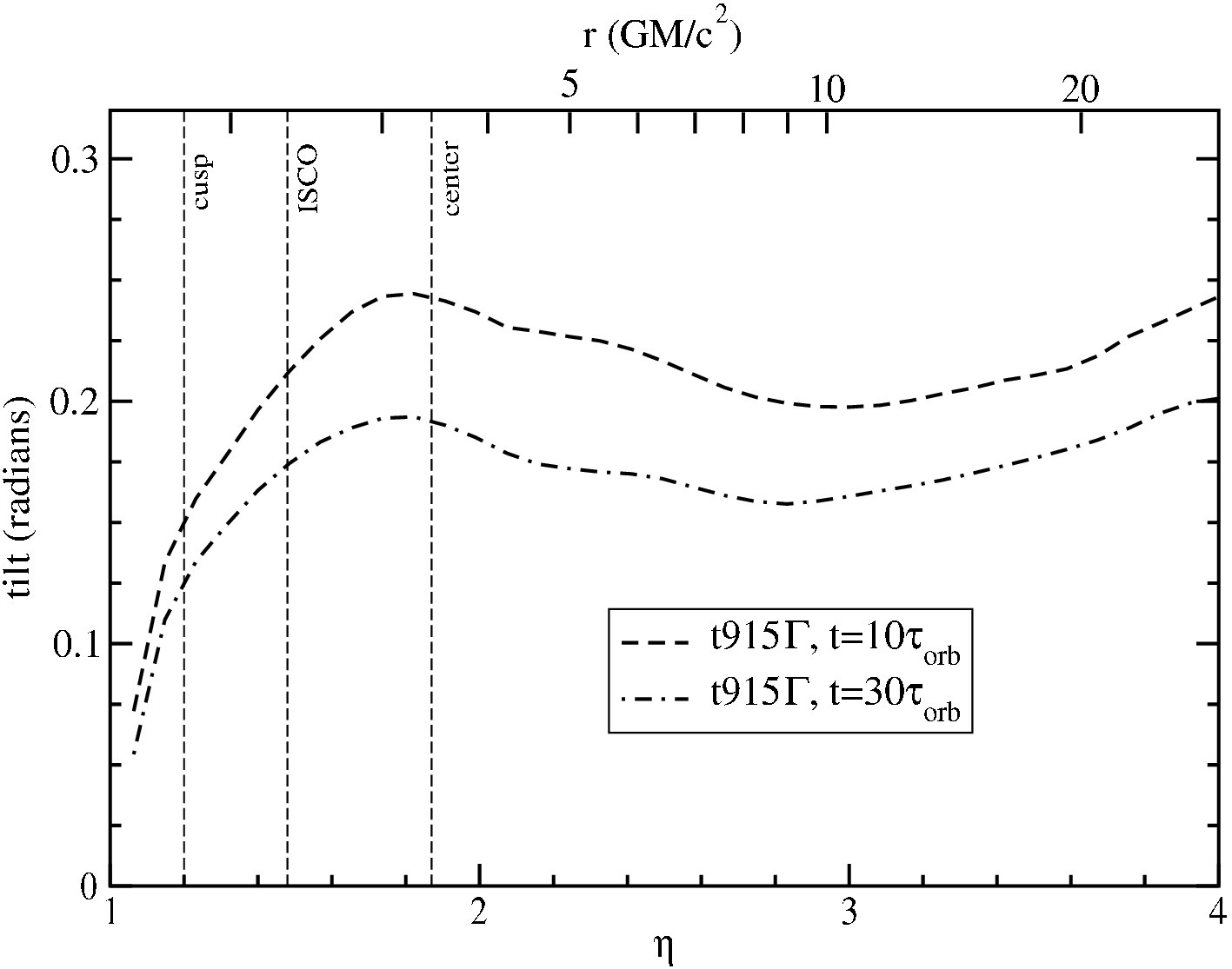}
\caption{}
\end{figure}

\begin{figure}
\figurenum{6d}
\plotone{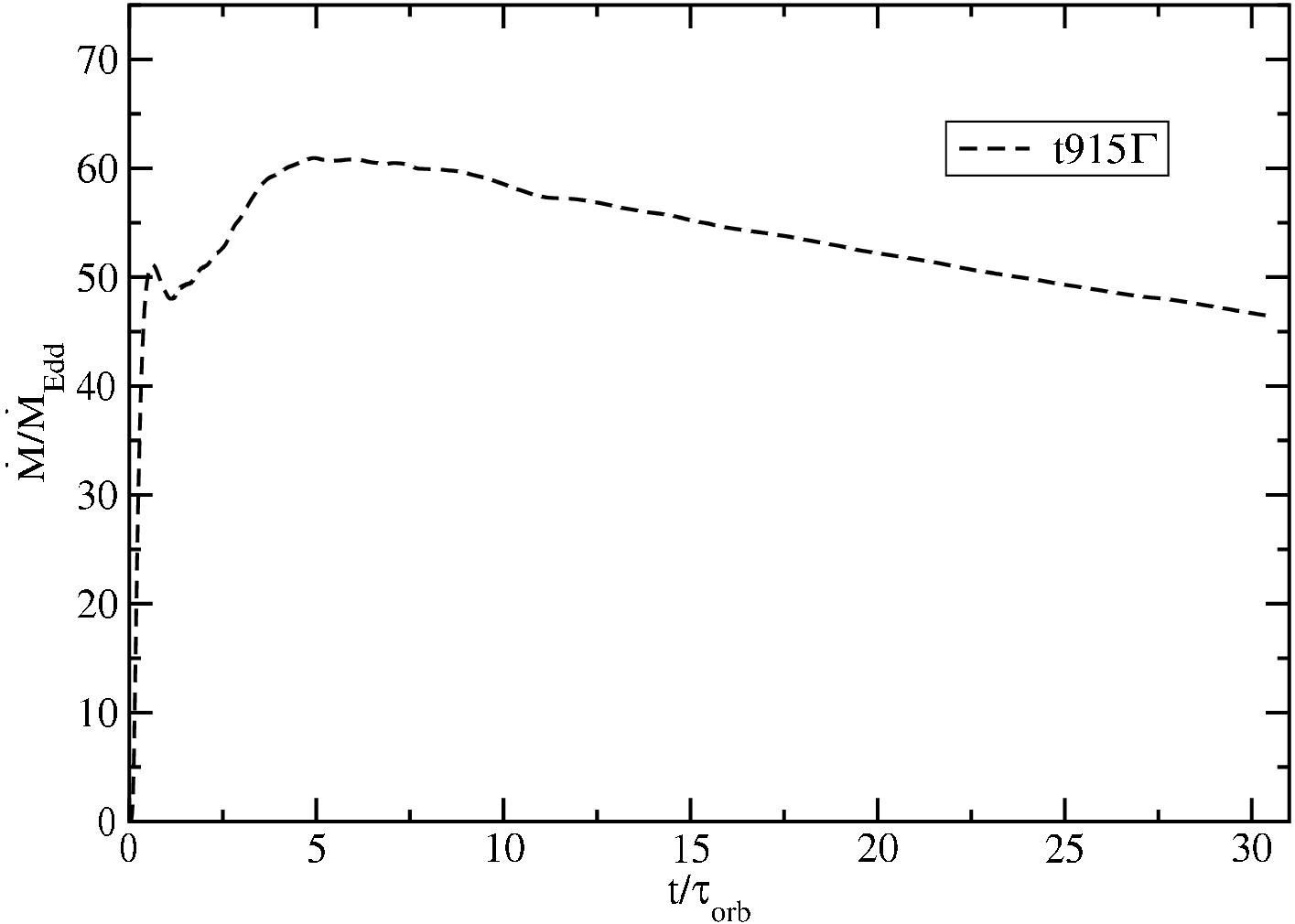}
\caption{}
\end{figure}

\begin{figure}
\figurenum{6e}
\plotone{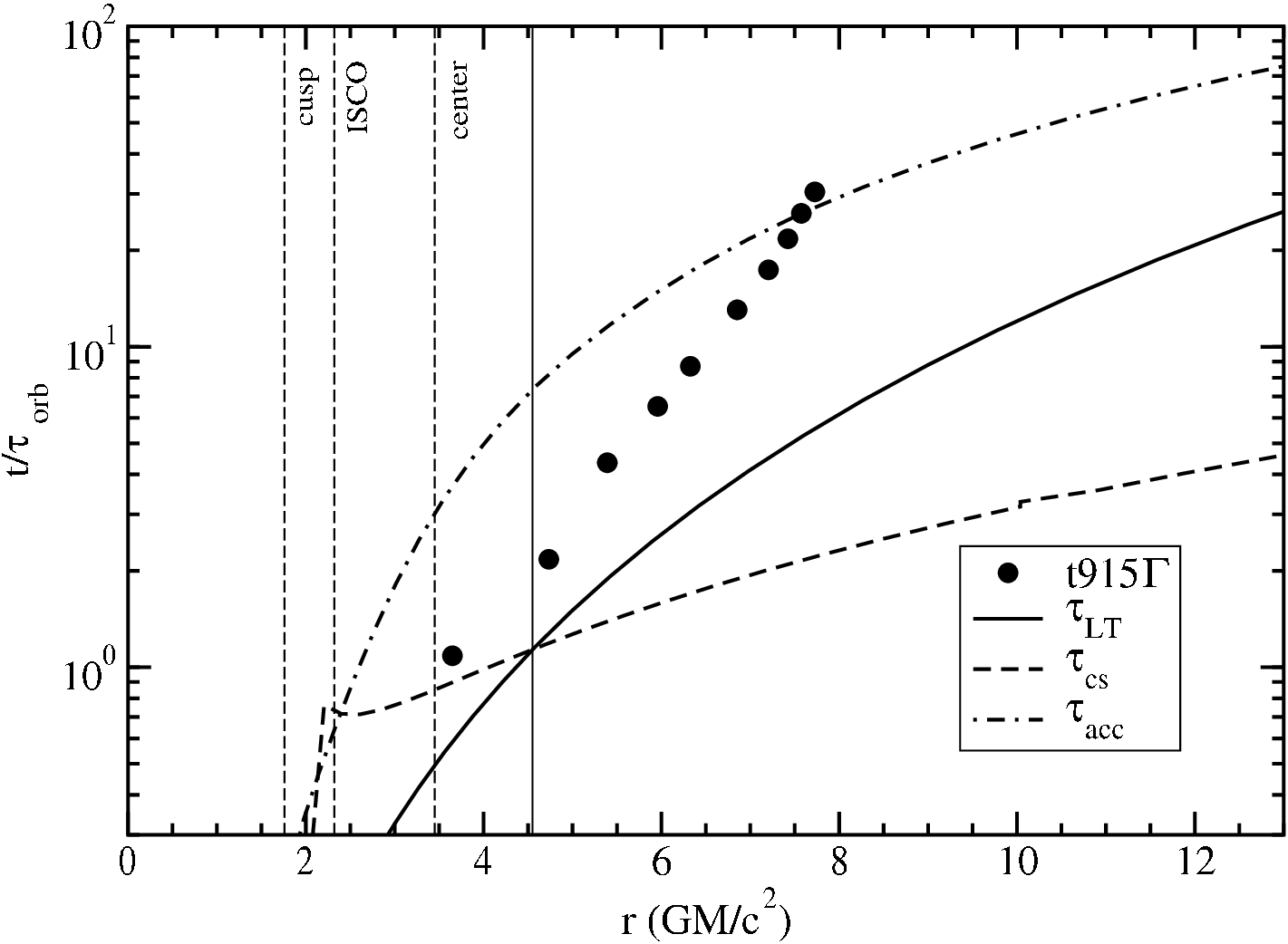}
\caption{}
\end{figure}


\subsection{Model t915R}
Model t915R explores a retrograde torus model.  The magnitude of the
black hole spin ($|a|=0.9$) is the same as in the previous models, but
it spins in the opposite sense.  Thus, the angular momentum of the black hole
and torus are almost anti-parallel; they differ from anti-parallel by
$15^\circ$.
The retrograde orbit requires the torus to have a higher specific angular
momentum to ensure $l>l_{ms}$.  In this case, $l/M=4.4751$; the torus
center is at $r_{center}/r_G=12$.
The cusp in the potential is located at
$r_{cusp}/r_G=6.67$, $\theta=\pi/2$ with $\Phi_{cusp}=-0.03$.
We choose an inner surface potential $\Phi_{in}=-0.0125$ giving
an energy gap $\Delta \Phi = 0.0175$.  The outer boundary is set
at $r_{max}/r_G=80$ and the simulation is evolved until
$t=2000 M\approx8\tau_{orb}$.

\begin{figure}
\figurenum{7a}
\plotone{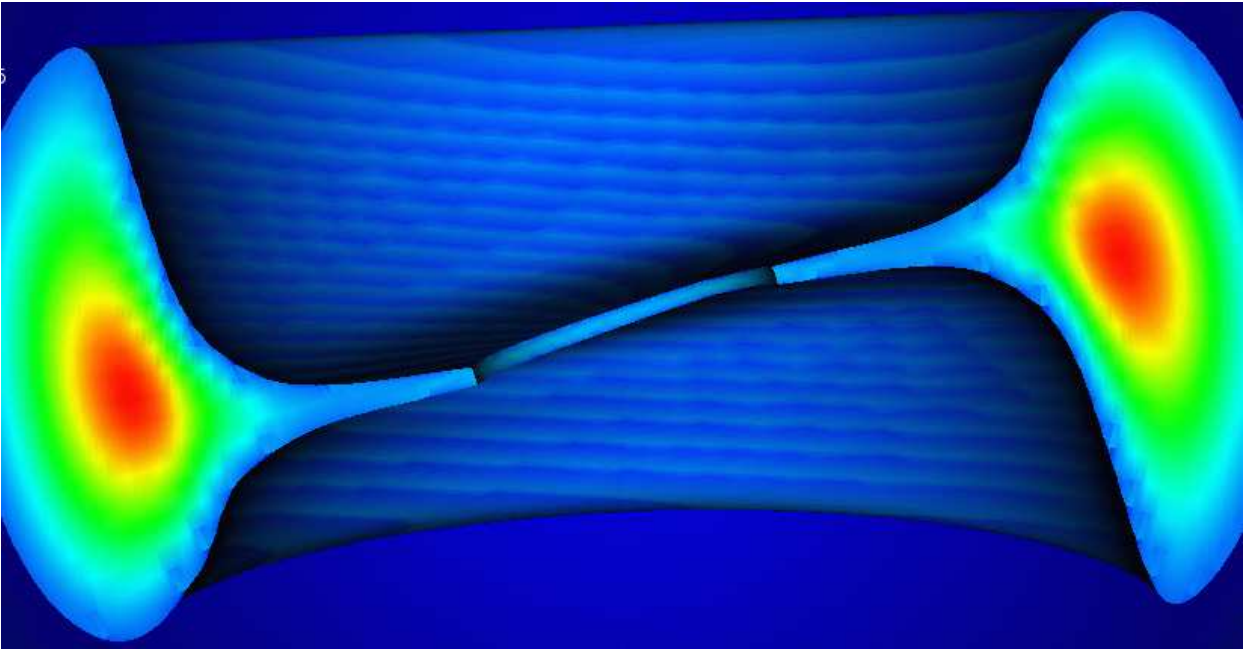}
\caption{Model t915R.
({\it a}) Linear density profile (similar to Figure \ref{fig:modelt90}{\it a})
at $t=2000M\approx8\tau_{orb}$.  The black hole spin axis is
tilted $15^\circ$ to the right of the $-z$-axis (points down and to
the right).
({\it b}) Cumulative
precession angle $\gamma$ and ({\it c}) tilt $\beta$ in radians as functions
of radius.  Initially, $\gamma=0$ and
$\beta=165^\circ\approx2.88$ throughout the disks.  ({\it d}) Mass accretion
rate in units of the Eddington mass accretion rate.  ({\it e}) Transition
radius ({\it symbols}) as a function of time along with the Lense-Thirring
precession timescale $\tau_{LT}$, azimuthal sound crossing time
$\tau_{cs}$, and accretion timescale
$\tau_{acc}$.  Here the thin vertical line marks
$r=r_{cusp}$.}
\label{fig:modelt915R}
\end{figure}

The most obvious thing we notice in Figure \ref{fig:modelt915R}{\it a} 
is that the accretion stream in this model takes the 
form of a relatively thin disk 
lying very close to the symmetry plane of the black hole.  
In this sense, the results appear very similar to the usual 
Bardeen-Petterson configuration, but for a very different physical 
reason as we shall see.  
The gas in the accretion stream still orbits in a retrograde sense 
just as the rest of the disk does.  
Therefore, the gas in the accretion stream has a tilt $\beta \approx \pi$ 
(Figure \ref{fig:modelt915R}{\it c}) consistent with retrograde 
motion near the symmetry plane of the black hole. Closest
to the black hole ($\eta<1.6$; $r/r_G<2.62$), the accreting gas
actually appears to tilt away from the symmetry plane. This is due 
the inclusion of background gas in our calculations.  
Close to the black hole, the background gas is undergoing very rapid 
prograde precession near the symmetry plane of the black hole.  
It, therefore, provides a $\beta \approx 0$ contribution near the 
black hole.

\begin{figure}
\figurenum{7b}
\plotone{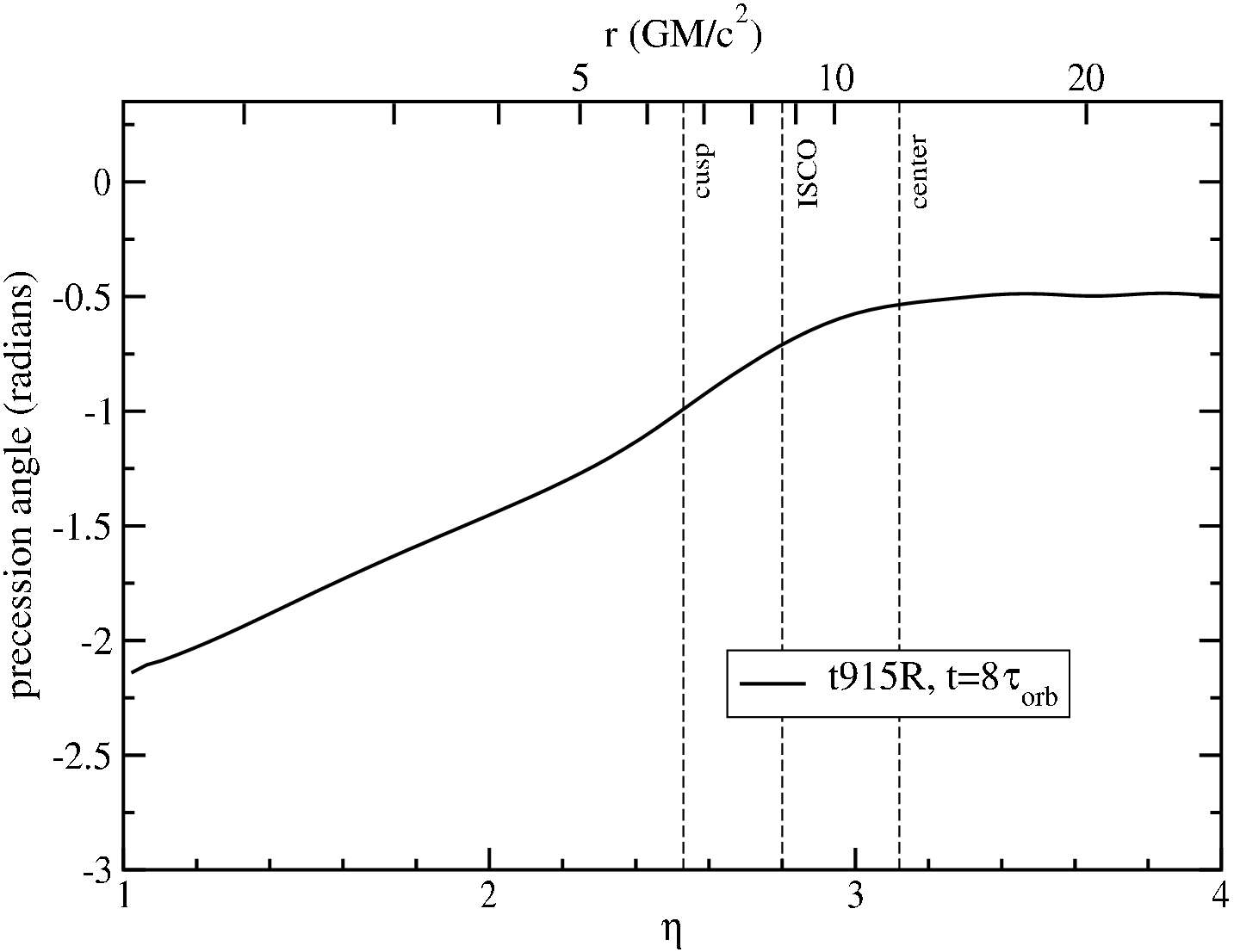}
\caption{}
\end{figure}

\begin{figure}
\figurenum{7c}
\plotone{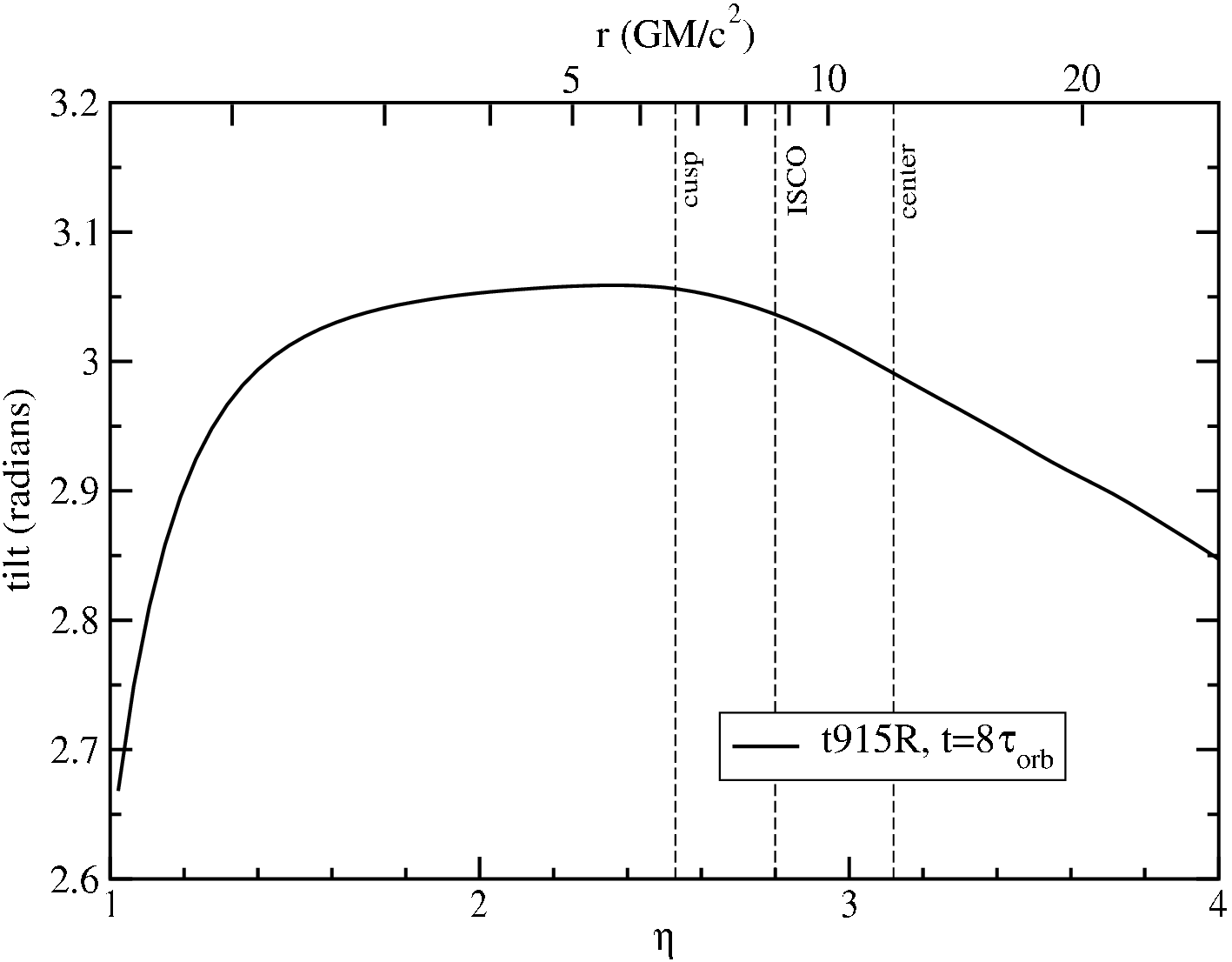}
\caption{}
\end{figure}

\begin{figure}
\figurenum{7d}
\plotone{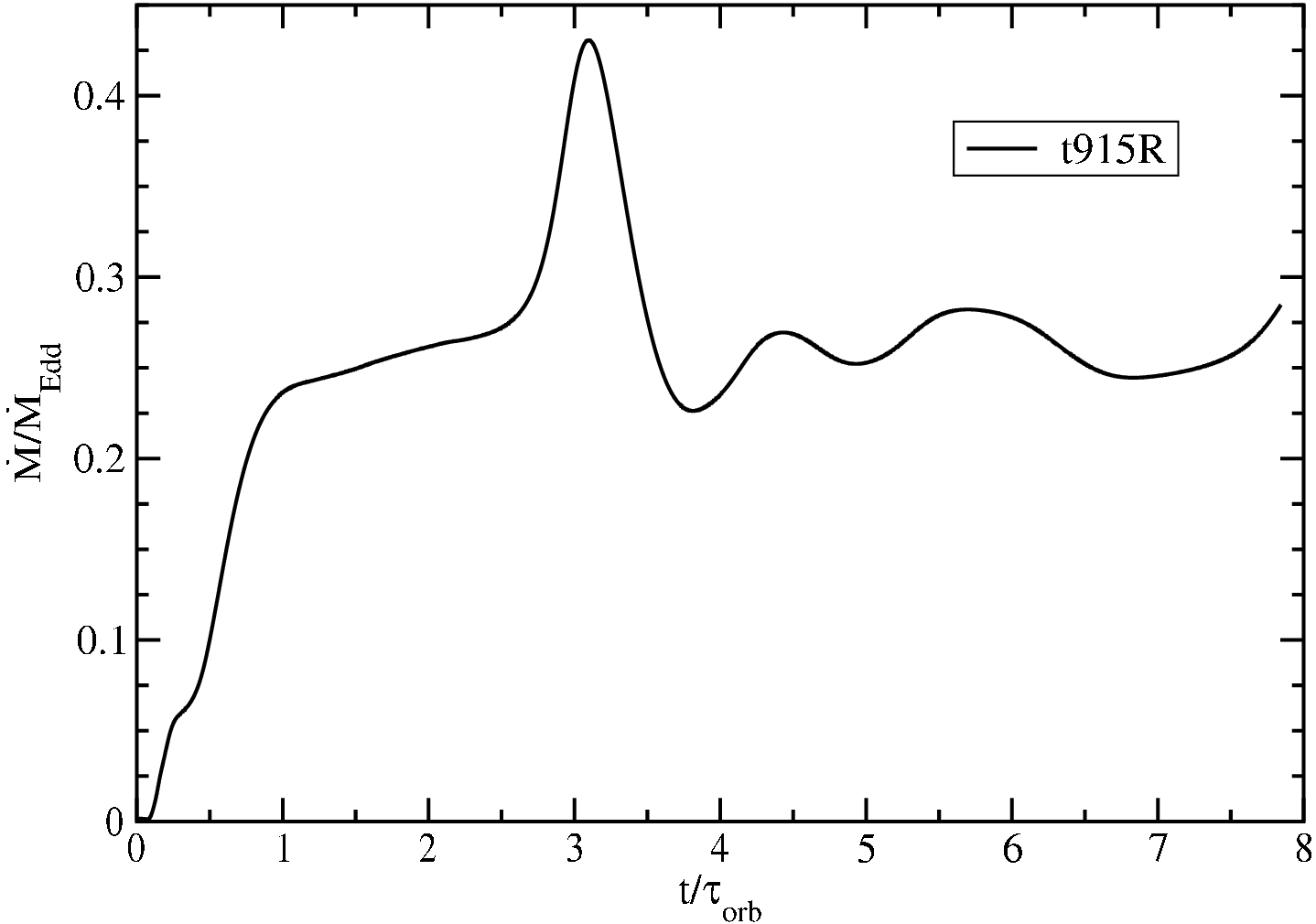}
\caption{}
\end{figure}

\begin{figure}
\figurenum{7e}
\plotone{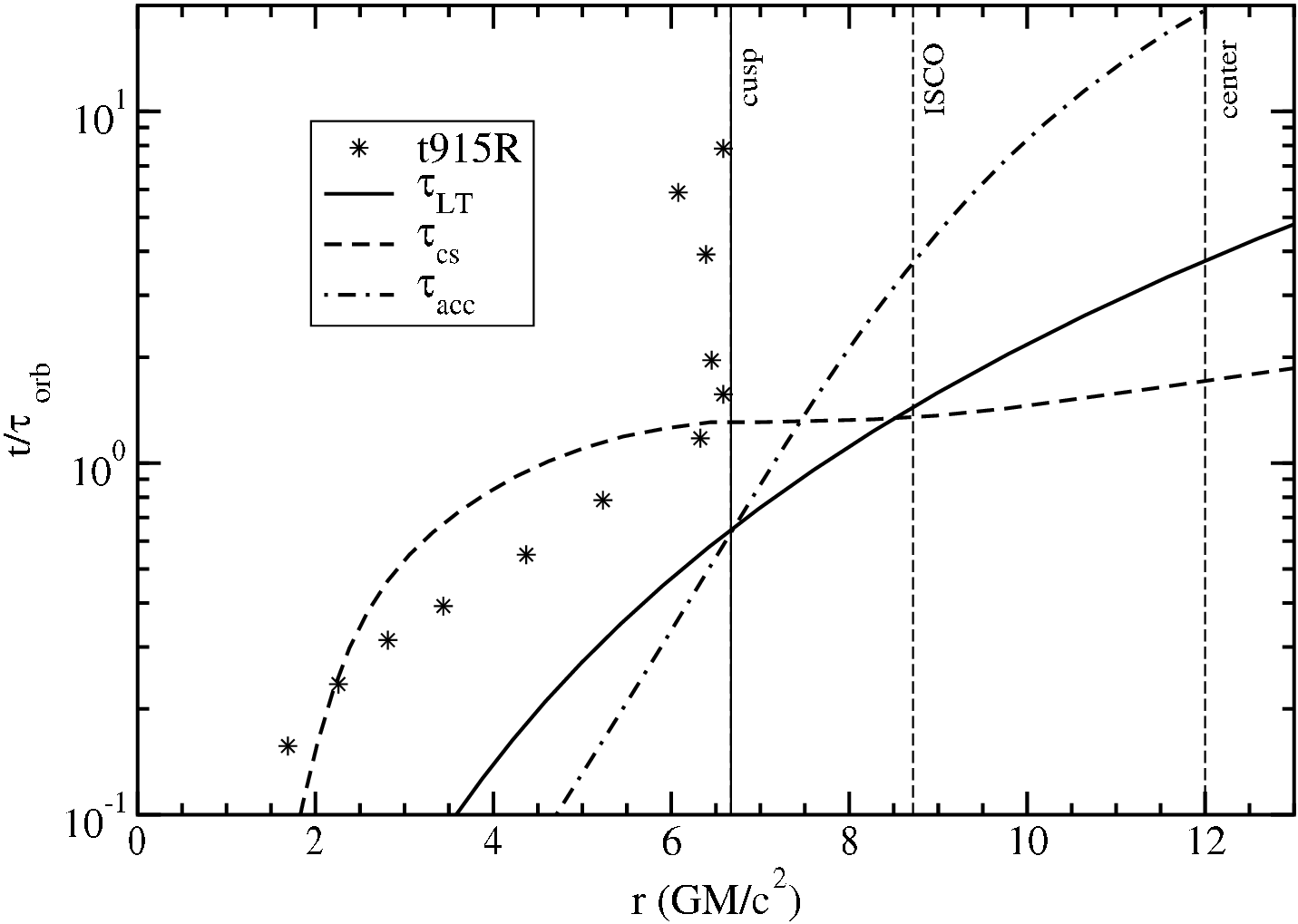}
\caption{}
\end{figure}

It is important to note that, 
within the accretion stream at $r < r_{cusp}$, 
the accretion timescale $\tau_{acc}$ is 
actually shorter than both the sound-crossing time $\tau_{cs}$ 
and the precession timescale $\tau_{LT}$ 
(Figure \ref{fig:modelt915R}{\it e}).  
We, therefore, don't expect 
precession to have a dramatic effect at these small radii 
in this model since the Lense-Thirring precession timescale is longer than the 
dynamical timescale.  The role of Lense-Thirring precession inside of 
$r_{cusp}$ is further diminished by the fact that most of the gas 
inside this radius lies very close to the symmetry plane of the black hole 
making precession essentially meaningless.  Notice, for instance, that 
the calculation of the precession 
$\gamma$, which relies on the cross-product of
$\mathbf{J}_{BH}$ and $\mathbf{J}_{Disk}$, which becomes unreliable whenever 
$\mathbf{J}_{BH}$ and $\mathbf{J}_{Disk}$ are nearly
parallel or anti-parallel.

Outside of $r_{cusp}$, the accretion timescale becomes significantly 
longer, yet the sound-crossing time quickly becomes shorter 
than the precession timescale.  Therefore, true differential precession 
is only expected to be important over a very small region in this disk 
model.  Importantly though, 
we see in Figure \ref{fig:modelt915R}{\it b} that the
precession proceeds in a retrograde fashion as
expected (Lense-Thirring precession always proceeds in the direction
of the black-hole spin).  
Precession also extends out to much larger radii (essentially 
throughout the disk) in this model than in previous models (see Figure 
\ref{fig:modelt915R}{\it b}).  
We will return to this point in the next section.

\subsection{Model t515L \& t515H}
Model t515 explores the dependence of our results on the magnitude of
the black hole angular momentum.  For these models, the black
hole has a spin $a/M=0.5$ and radius $r_{BH}/r_G=1.87$;
the torus has a specific angular
momentum $l/M=3.385$; the torus center is at $r_{center}/r_G=6.92$.
The cusp in the potential is located at
$r_{cusp}/r_G=2.96$, $\theta=\pi/2$ with $\Phi_{cusp}=-0.0136$.
We choose an inner surface potential $\Phi_{in}=-0.0096$ giving
an energy gap $\Delta \Phi = 0.004$.  The outer boundary is set
at $r_{max}/r_G=120$.
Model t515L uses a resolution of $48\times24\times48$ zones and is evolved
until $t=2400 M\approx20\tau_{orb}$;
Model t515H uses a resolution of $96\times64\times96$
and is evolved until $t=1200 M\approx10\tau_{orb}$.
Since the grid in these models has a larger maximum radius
than most of the other models, the effective resolution
is somewhat less.

\begin{figure}
\figurenum{8a}
\plotone{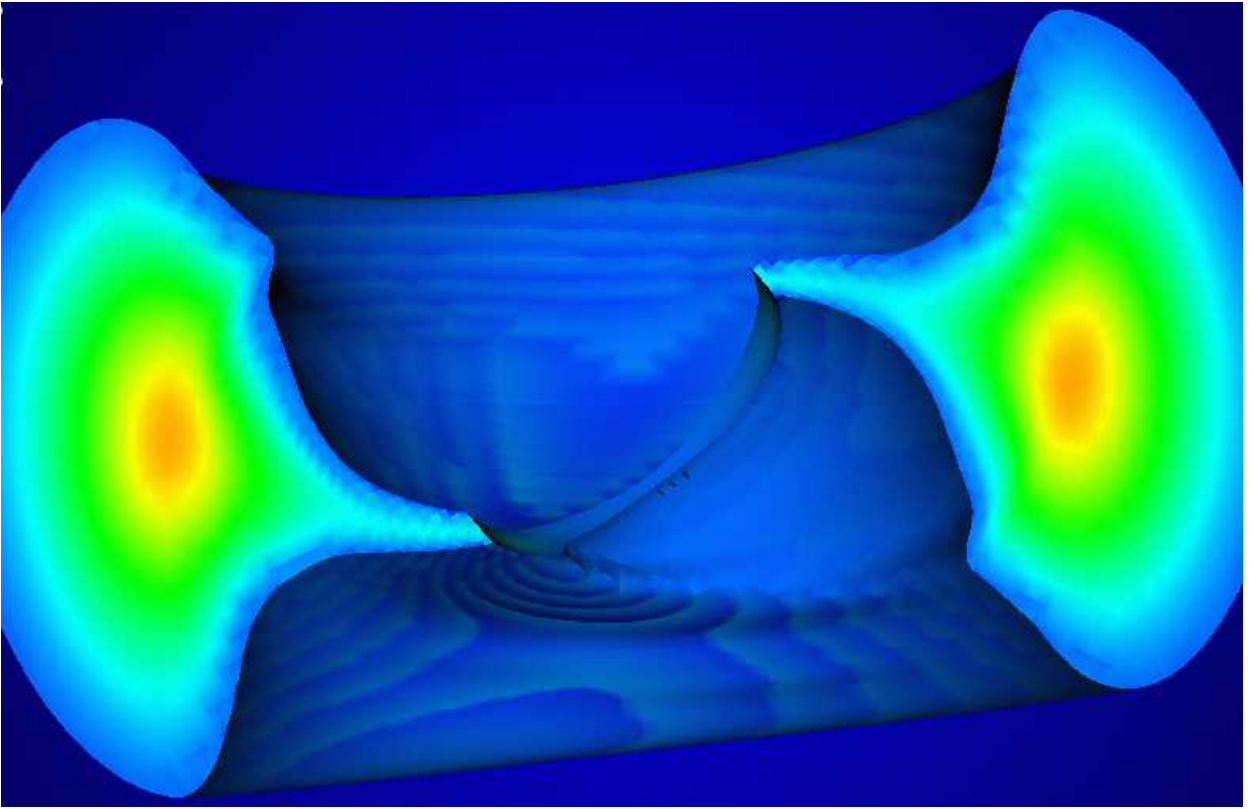}
\caption{Model t515.
({\it a}) Linear density profile (similar to Figure \ref{fig:modelt90}{\it a})
at $t=1200M\approx10\tau_{orb}$.  The black hole spin axis is
tilted $15^\circ$ to the left of the $z$-axis.
({\it b}) Cumulative
precession angle $\gamma$ and ({\it c}) tilt $\beta$ in radians as functions
of radius.  Initially, $\gamma=0$ and
$\beta=15^\circ\approx0.26$ throughout the disks.  In panel {\it b}, 
we have included data from $t=0.3$, 1, 1.6, 3.2, 6.4, and $8 \tau_{orb}$ 
({\it dotted lines}) to give 
a more complete time history of the precession.  ({\it d}) Mass accretion
rate in units of the Eddington mass accretion rate.  ({\it e}) Transition
radius ({\it symbols}) as a function of time along with the Lense-Thirring
precession timescale $\tau_{LT}$, azimuthal sound crossing time
$\tau_{cs}$, and accretion timescale
$\tau_{acc}$.  The thin vertical line marks the radius
where $\tau_{LT}=\tau_{cs}$.}
\label{fig:modelt515}
\end{figure}

There are two striking features to these models. First, in Figures
\ref{fig:modelt515}{\it a} and {\it c}, we see that the inner
accretion disk appears to tilt {\it away} from the symmetry
plane of the black hole. This is actually consistent with what we have seen
in previous models in the sense that the black hole preferentially
accretes the most tilted material.  As in previous models, two
opposing accretion streams begin from latitudes well above and below 
the black-hole symmetry plane near
$r_{center}$. The accretion streams in this model actually begin
at comparable latitudes to what is seen in Model t915 (compare Figures
\ref{fig:modelt515}{\it a} and \ref{fig:modelt915}{\it a}). Again, this
material carries much of the tilted angular momentum into the hole
and allows the disk to gradually align. Since the streams have a
greater tilt on average than the material further out,
$\beta$ increases toward smaller radii inside $r_{center}$ as seen in Figure
\ref{fig:modelt515}{\it c}. However, unlike our $a/M=0.9$ models, these
accretion streams do not cross before passing into the
hole and we do not see the 
formation of an inner ring of low $\beta$ material.  
Consequently, $\beta$ remains large from near $r_{center}$ 
all the way in to the horizon.

\begin{figure}
\figurenum{8b}
\plotone{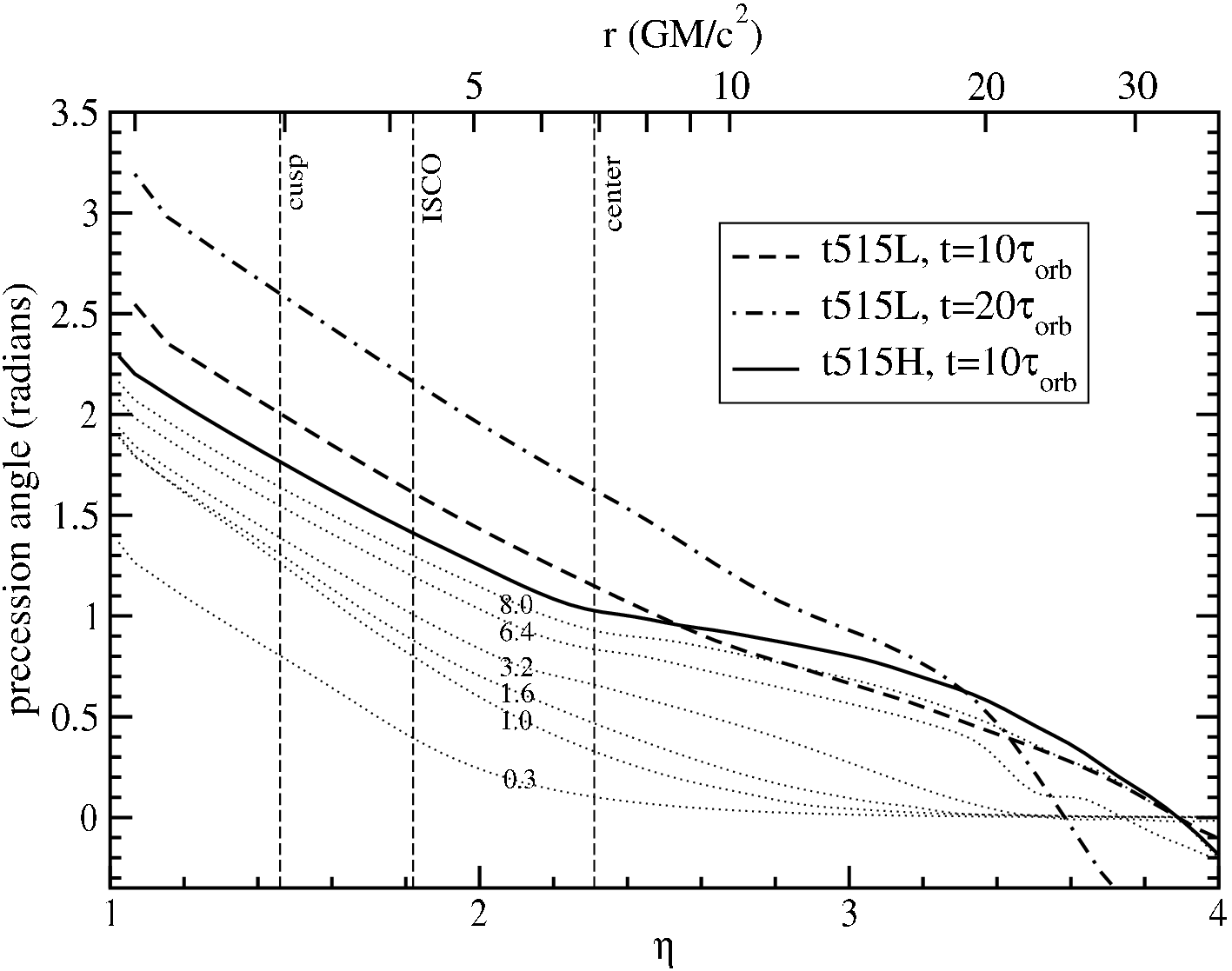}
\caption{}
\end{figure}

\begin{figure}
\figurenum{8c}
\plotone{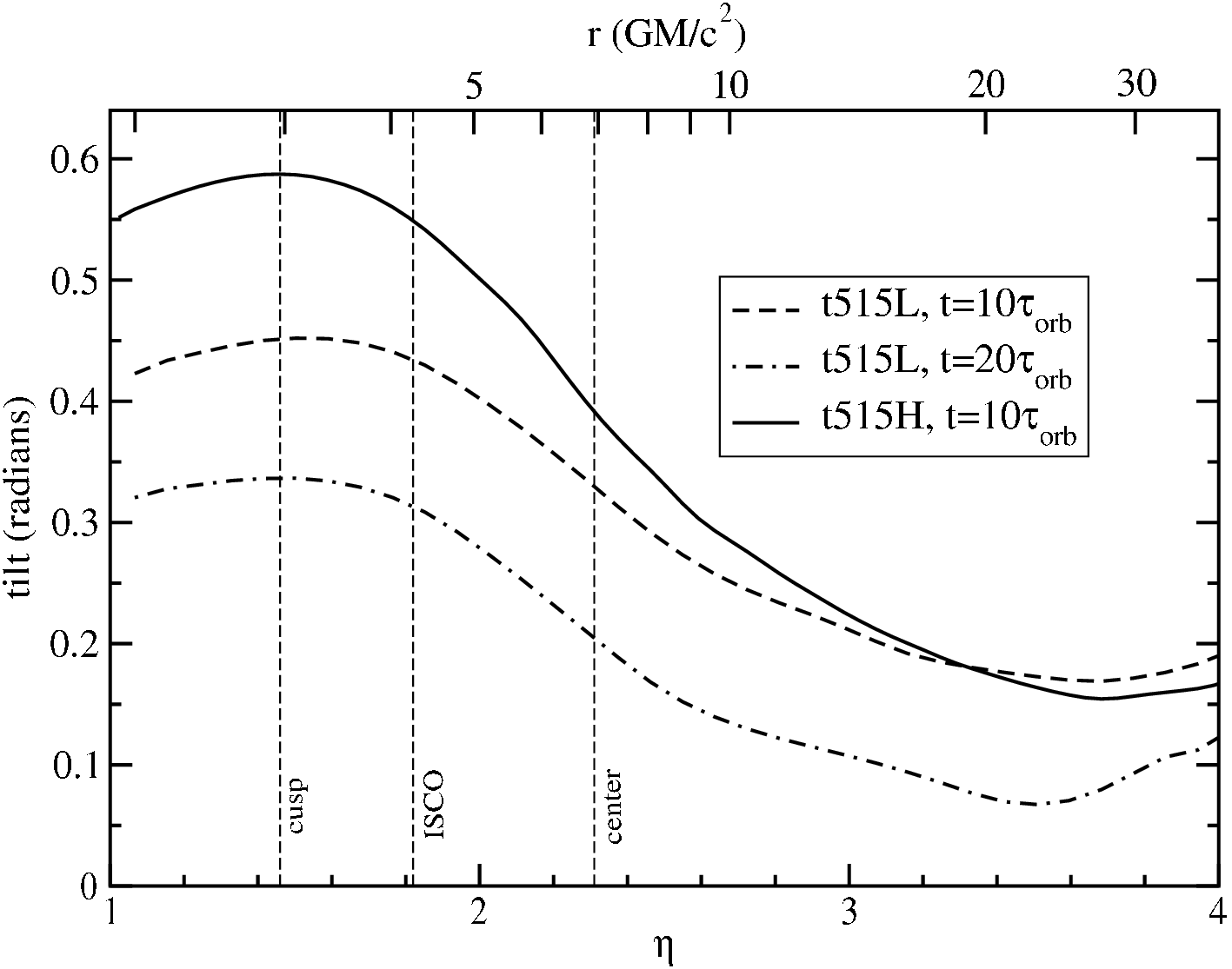}
\caption{}
\end{figure}

\begin{figure}
\figurenum{8d}
\plotone{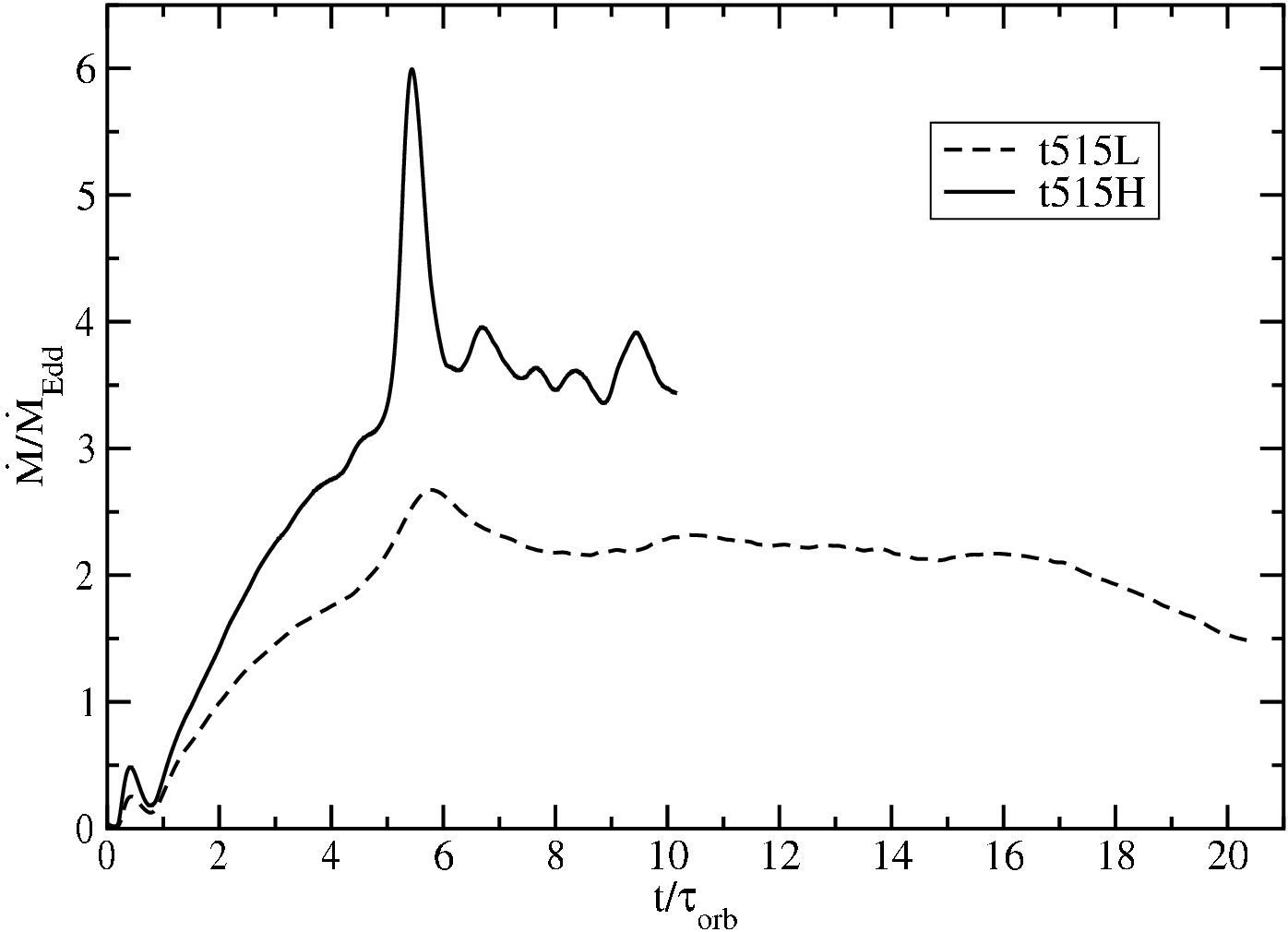}
\caption{}
\end{figure}

\begin{figure}
\figurenum{8e}
\plotone{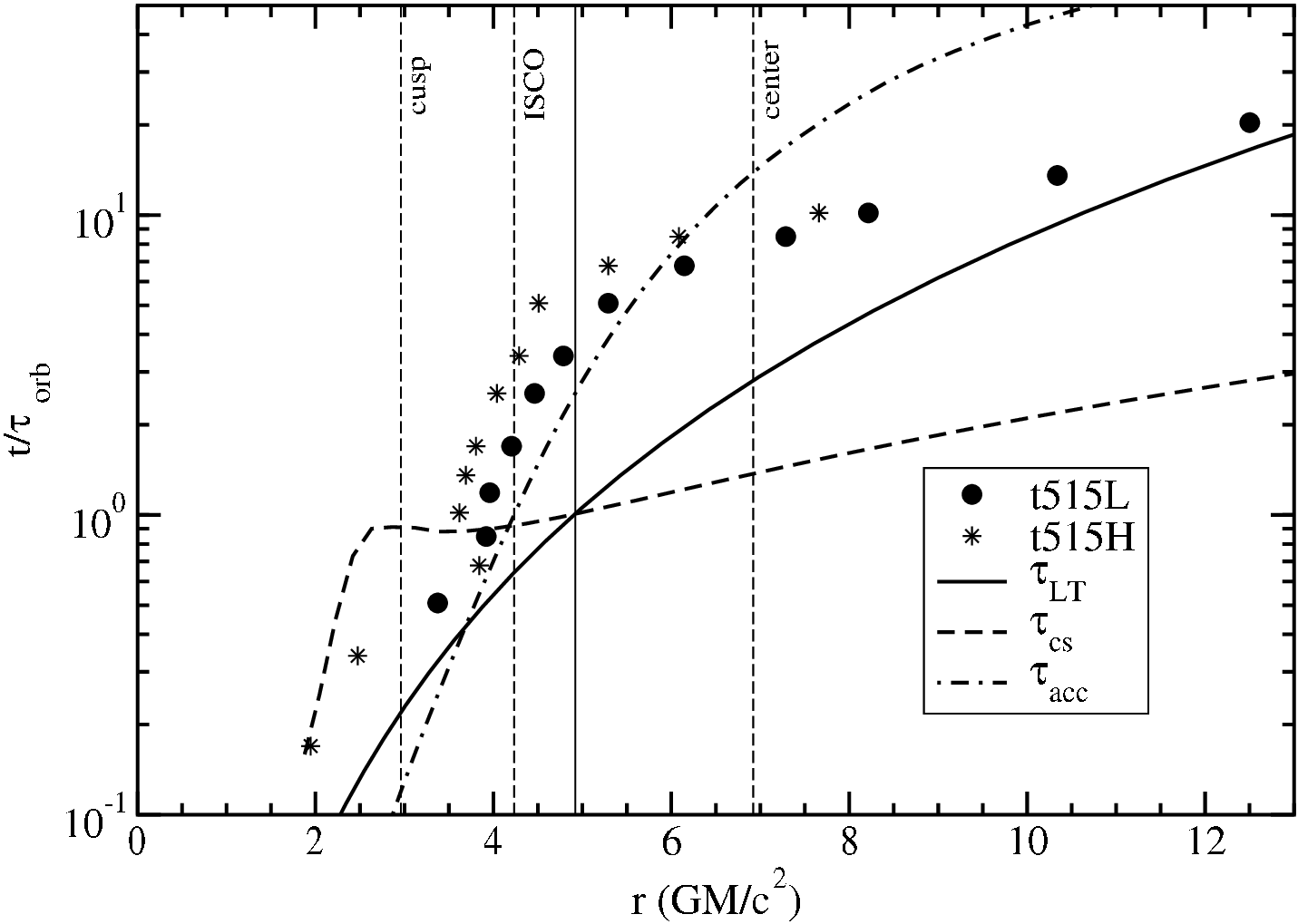}
\caption{}
\end{figure}

The second striking feature is apparent in Figure
\ref{fig:modelt515}{\it e}.  Although the transition radius at first
approaches an asymptotic value coincident with
$\tau_{LT}\approx\tau_{cs}$ similar to our other models, it later
appears to rapidly move outward with no sign of another limit being
reached. This is actually evidence of near rigid-body precession in the disk. 
For pure rigid-body precession, the twist $\gamma$ in Figure 
\ref{fig:modelt515}{\it b} would be a horizontal line and the transition 
radius would lose its meaning since there would be no ``transition'' from 
an inner, twisted region to an outer, untwisted one.
Model t515 doesn't quite undergo pure rigid-body precession.  
It actually starts out undergoing differential precession and then 
switches to near rigid-body precession.  This change is most apparent in the
time sequence of $\gamma$-profiles shown in Figure
\ref{fig:modelt515}{\it b}.  This also explains the near
constant precession angle for gas beyond $r=r_{center}$ in Figures
\ref{fig:modelt915R}{\it b}.

Generally, rigid-body precession is 
expected in disks whenever $\tau_{cs}$ is less than the
applicable precession timescale \citep{lar96}, in this case whenever
$\tau_{cs}<\tau_{LT}$. This condition is actually met in all of our
models at large enough radii; however, only in Models t515 and t915R is
this condition met for $r<r_{center}$. Thus, only in these two models
does the bulk of the gas in the disk meet the condition for rigid-body 
precession.

\section{Discussion and Summary}
\label{sec:discussion}

We have performed the first fully relativistic numerical studies of
tilted accretion disks around Kerr black holes.  Such disks are
subject to Lense-Thirring precession, resulting in a
torque that tends to twist and warp the disk. For the thick disks
considered here, we find that the nature of this precession depends 
primarily on the sound speed in the disk.  Whenever $\tau_{LT}<\tau_{cs}$ 
in the bulk of the disk, it undergoes differential precession out to 
a transition radius set by $\tau_{LT} \approx \tau_{cs}
\lesssim \tau_{acc}$, which occurs at a relatively modest distance
of $r/r_G \approx 7$.  In other disk models, such as viscous
Keplerian disks, this transition may occur at somewhat larger radii
\citep[$\sim15-30 r_G$,][]{nel00}. We find that the
location of this transition radius does not depend strongly on the
size, mass, mass accretion rate, or tilt of the disk. The initial
tilt does, however, affect the overall mass accretion rate for our
models, since higher tilt models begin further from equilibrium within
the black hole potential. 

Whenever $\tau_{LT}>\tau_{cs}$ for the bulk of the disk, we
find that it undergoes near rigid-body precession after a
short initial period of differential precession. Such rigid-body
precession of a thick disk could have important astrophysical
consequences. For instance, this is the basis of at least one model
attempting to explain the 106 day variability observed from 
Sgr A* \citep{liu02}, the $2.6\times10^6 M_\odot$ black hole 
at the Galactic center. It could
also be important for understanding the jet precession observed 
in Galactic microquasars such as SS 433 and Her X-1.


One obvious shortcoming of this work is that 
we have ignored angular momentum transport mechanisms
that are likely present in most astrophysical disks [e.g. viscosity
or the magneto-rotational instability \citep{bal91}]. Angular
momentum dissipation may allow the inner region of the tilted disk
to align more quickly with the symmetry plane of the black hole than
what we see in the inviscid disks considered here
\citep{bar75,kum85,sch96}. The alignment may also be more efficient
in viscous disks, resulting in a sharper transition \citep{nel00}. A
sharp transition from an inner aligned disk to a tilted outer disk
may be useful in explaining a variety of phenomena observed in
black-hole systems, such as quasi-periodic oscillations
\citep{fra01a} as well as certain spectral features.



\begin{acknowledgements}
The authors would like to thank the VisIt development team at
Lawrence Livermore National Laboratory (http://www.llnl.gov/visit/),
in particular Hank Childs, for visualization support.
P.C.F. would like to acknowledge many useful discussions on this
work, particularly with G. Mathews, J. Wilson, and C. Gammie.
This work was performed
under the auspices of the U.S. Department of Energy by
University of California, Lawrence
Livermore National Laboratory under Contract W-7405-Eng-48.
\end{acknowledgements}

\clearpage
\bibliographystyle{apj}
\bibliography{myrefs}

\end{document}